\documentclass[twocolumn]{aastex62}
\usepackage{graphicx}
\usepackage{color}
\usepackage{amsmath,amssymb}
\usepackage{times}
\begin{document}

\correspondingauthor{Emily C. Cunningham}
    \email{eccunnin@ucsc.edu}
	
	\author[0000-0002-6993-0826]{Emily C. Cunningham}
	\affiliation{Department of Astronomy \& Astrophysics,
    			University of California, Santa Cruz,
				1156 High Street,
				Santa Cruz, CA 95064, USA}
	\title{HALO7D II: The Halo Velocity Ellipsoid and Velocity Anisotropy with Distant Main Sequence Stars}
    
    \author[0000-0001-6146-2645]{Alis J. Deason}
    \affiliation{Institute for Computational Cosmology, Department of Physics, University of Durham, South Road, Durham DH1 3LE, UK}
    
    \author[0000-0003-3939-3297]{Robyn E. Sanderson}
	\affiliation{Department of Physics \& Astronomy, University of Pennsylvania, 209 S 33rd St., Philadelphia, PA 19104, USA}
	\affiliation{Center for Computational Astrophysics, Flatiron Institute, 162 5th Ave., New York, NY 10010, USA}
    
    \author[0000-0001-8368-0221]{Sangmo Tony Sohn}
    \affiliation{Space Telescope Science Institute, 3700 San Martin Drive, Baltimore, MD 21218, USA}
    
    \author[0000-0003-2861-3995]{Jay Anderson}
    \affiliation{Space Telescope Science Institute, 3700 San Martin Drive, Baltimore, MD 21218, USA}
    
    \author[0000-0001-8867-4234]{Puragra Guhathakurta}
    \affiliation{Department of Astronomy \& Astrophysics,
    			University of California, Santa Cruz,
				1156 High Street,
				Santa Cruz, CA 95064, USA}
    
    \author{Constance M. Rockosi}
    \affiliation{Department of Astronomy \& Astrophysics,
    			University of California, Santa Cruz,
				1156 High Street,
				Santa Cruz, CA 95064, USA}
    
    \author[0000-0001-7827-7825]{Roeland P. van der Marel}
    \affiliation{Space Telescope Science Institute, 3700 San Martin Drive, Baltimore, MD 21218, USA}
    \affiliation{Center for Astrophysical Sciences, Department of Physics \& Astronomy, Johns Hopkins University, Baltimore, MD 21218, USA}
    
    \author[0000-0003-3217-5967]{Sarah R. Loebman}
    \affiliation{Department of Physics, University of California, Davis, CA 95616, USA}
    
    \author[0000-0003-0603-8942]{Andrew Wetzel}
    \affiliation{Department of Physics, University of California, Davis, CA 95616, USA}

	\begin{abstract}
		The Halo Assembly in Lambda-CDM: Observations in 7 Dimensions (HALO7D) dataset consists of Keck II/DEIMOS spectroscopy and Hubble Space Telescope-measured proper motions of Milky Way (MW) halo main sequence turnoff stars in the CANDELS fields. In this paper, the second in the HALO7D series, we present the proper motions for the HALO7D sample. We discuss our measurement methodology, which makes use of a Bayesian mixture modeling approach for creating the stationary reference frame of distant galaxies. Using the 3D kinematic HALO7D sample, we estimate the parameters of the halo velocity ellipsoid, $\langle v_{\phi} \rangle, \sigma_r, \sigma_{\phi}, \sigma_{\theta}$, and the velocity anisotropy $\beta$. Using the full HALO7D sample, we find $\beta=0.63 \pm 0.05$ at $\langle r \rangle =24$ kpc. We also estimate the ellipsoid parameters for our sample split into three apparent magnitude bins; the posterior medians for these estimates of $\beta$, while consistent with one another, increase as a function of mean sample distance. Finally, we estimate $\beta$ in each of the individual HALO7D fields. We find that the velocity anisotropy $\beta$ can vary from field to field, which suggests that the halo is not phase mixed at $\langle r \rangle =24$ kpc. We explore the $\beta$ variation across the skies of two stellar halos from the \textit{Latte} suite of FIRE-2 simulations, finding that both simulated galaxies show $\beta$ variation over a similar range to the variation observed across the four HALO7D fields. The accretion histories of the two simulated galaxies result in different $\beta$ variation patterns; spatially mapping $\beta$ is thus a way forward in characterizing the accretion history of the Galaxy.
	\end{abstract}
    
	
	\section{Introduction}    
    
    The Milky Way (MW) stellar halo's kinematic structure contains key clues about the Galaxy's formation and mass assembly. According to the Lambda Cold Dark Matter ($\Lambda$CDM) paradigm for the evolution of the universe, the MW has built up its halo of dark matter over cosmic time by accreting smaller dark matter halos, some of which host dwarf galaxies. The remnants of these accreted dwarfs are found in the Milky Way's stellar halo, and the velocities of these stars retain a link to their initial conditions because of their long dynamical times. The HALO7D project aims to investigate the MW's formation by studying the chemical and phase-space structure of the stellar halo's distant, main sequence (MS) stars. 
    
    One kinematic quantity that has long been of interest in MW formation studies is the velocity anisotropy $\beta$ (\citealt{Binney2008}), which provides a measure of the relative energy in tangential and radial orbits: 
    
    \begin{equation}
    \beta = 1 - \frac{\langle v_{\phi}^2 \rangle +\langle v_{\theta}^2 \rangle}{2 \langle v_r^2 \rangle }.
    \end{equation}
Systems with $\beta=1$ are on completely radial orbits, while a population of stars on perfectly circular orbits has $\beta=-\infty$. 

The velocity anisotropy parameter $\beta$ plays a key role in the spherical \cite{Jeans1915} equation:

\begin{equation}
    M_{\rm Jeans}(<r) = -\frac{r\sigma_r^2}{G}\left(\frac{\mathrm{d} \ln \rho}{\mathrm{d} \ln r} + \frac{\mathrm{d} \ln \sigma_r^2}{\mathrm{d} \ln r}+2\beta\right).
    \label{eqn:jeans}
    \end{equation}
Jeans modeling has been used to estimate the mass of the Galaxy in many studies (e.g., \citealt{Dehnen2006}, \citealt{Gnedin2010}, \citealt{Watkins2009}, \citealt{Deason2012}, \citealt{Eadie2017}, \citealt{Sohn2018}, \citealt{Watkins2018} and references therein). However, estimates of the MW's mass have long been plagued by the mass-anisotropy degeneracy, owing to the lack of constraints on the tangential velocity distributions. 
It has only recently become possible to directly measure the tangential motion of kinematic tracers outside of the solar neighborhood. Previous studies have estimated $\beta$ from line-of-sight (LOS) velocities alone (e.g., \citealt{Sirko2004}; \citealt{Kafle2012}; \citealt{Deason2012}, \citealt{King2015}), taking advantage of the fact that, because of our position within the Galaxy, the LOS velocity distribution contains information about the tangential velocity distributions. However, as pointed out by \cite{Hattori2017}, studies of stars beyond $r \sim 15$ kpc with only LOS data (where $v_{LOS} \approx v_r$) result in systematic underestimates of $\beta$. 
        
    Fortunately, measuring tangential properties of tracers is now possible, thanks to the Hubble Space Telescope (\textit{HST}) and the \textit{Gaia} mission. The first estimate of $\beta$ outside the solar neighborhood using directly measured 3D kinematics was presented by \cite{Cunningham2016}, hereafter C16, using 13 MS stars with PMs measured from \textit{HST} and radial velocities measured from Keck spectra. We found $\beta =-0.3^{+0.4}_{-0.9}$, consistent with isotropy and lower than solar neighborhood estimates, which find a radially biased $\beta \sim 0.5-0.7$ (\citealt{Smith2009}, \citealt{Bond2010}). However, the uncertainties on this measurement were substantial (primarily due to the small sample size), and in order to better constrain $\beta$ and the MW mass, more tracers are required.
    
    Studies have recently used the PMs of globular clusters (GCs) as kinematic tracers to estimate $\beta$ and the mass of the MW. \cite{Sohn2018} used their own \textit{HST} PM measurements of 16 GCs to find $\beta=0.609^{+0.130}_{-0.229}$ in the Galactocentric distance range of $R_{\rm GC} = 10$--40~kpc, and a corresponding MW virial mass of $M_{\rm MW, virial}=2.05^{0.97}_{-0.79} \times 10^{12} M_{\odot}$. \cite{Watkins2018} used PM determinations of 34 GCs in the range $R_{\rm GC} = 2.0$--21.1~kpc based on {\it Gaia} DR2 \citep{gaia_kinematics2018} and found $\beta=0.48^{+0.15}_{-0.20}$ consistent with \cite{Sohn2018}, and a corresponding virial mass of $M_{\rm MW, virial}=1.41^{1.99}_{-0.52} \times 10^{12} M_{\odot}$.
        
    While studies have sought to estimate a single value $\beta$ in order to estimate the mass of the MW, studies of $\beta$ can have additional power in constraining the MW's accretion history. For example, the anisotropy radial profile $\beta(r)$ can contain information about the Galaxy's assembly history. In \cite{Deason2013b} and C16, we argued that our isotropic measurement of $\beta$, which is lower than solar neighborhood measurements and also distant halo estimates (\citealt{Deason2012}), indicates a ``dip'' in the $\beta$ profile, and that this dip could indicate the presence of a shell. 
    
    \cite{Loebman2018} provided theoretical perspective on this question, by studying the $\beta$ profiles in three suites of simulations, including accretion-only and cosmological hydrodynamic simulations. They found that both types of simulations predict radially biased $\langle \beta \rangle \sim 0.7$ beyond 10 kpc. Only one of the 17 simulations studied had tangentially biased $\beta$ over a large range of radii at $z=0$; this extended $\beta$ dip was the result of a major merger at $z\sim1$. While the other 16 simulations had radially biased $\beta$ at $z=0$, \cite{Loebman2018} found that temporal dips in the $\beta$ profile could arise. They found that recently accreted material can result in short-lived dips in $\beta$, while the passage of a massive satellite can induce a longer-lived dip in the $\beta$ profile from the in-situ component of the stellar halo. This latter scenario could explain the observed ``dip'' along the line of sight towards M31, as recent studies of the Triangulum Andromeda overdensity have suggested that its origin may be the disk rather than an accreted satellite (\citealt{Price-Whelan2015}; \citealt{Bergemann2018}), and that the event that disturbed the orbits of these disk stars may be the passage of the Sagittarius dwarf (\citealt{Laporte2018}).
    
    The anisotropy variation across different subpopulations in the halo can also be used to disentangle accretion events. Using 7D measurements from the \textit{Gaia} DR1 and SDSS of local MS stars, \cite{Belokurov2018} found that the relatively metal-rich stars ($[\mathrm{Fe/H}]>-1.7$) show strongly radially biased velocity anisotropy (i.e., ``sausage" stars, named thus because of the elongated radial velocity distribution relative to the tangential velocity distribution), while the metal-poor stars display an isotropic velocity distribution. They argue that presence of this radially biased, relatively metal-rich population in the inner halo indicates that the MW experienced a relatively massive, early accretion event. Evidence for this scenario has been bolstered with results from \textit{Gaia} DR2 (\citealt{Helmi2018}, \citealt{Deason2018b}). \cite{Lancaster2018} showed that the kinematics of the BHBs in \textit{Gaia} DR2 can be modeled by a mixture of two populations: one strongly radially biased and one isotropic. Debris from a massive, radialized dwarf that dominates the inner halo, known as the \textit{Gaia}-Sausage, \textit{Gaia}-Enceladus, or Kraken, is speculated to be responsible for this signature.
    
    Thanks to the \textit{Gaia} mission, it is now possible to estimate the $\beta$ of stars in the MW; however, even with {\it Gaia} DR2, uncertainties remain substantial at large radii, and, even in the the final data release, \textit{Gaia} will provide PMs only for stars brighter than $G\sim 20$. As a result, \textit{Gaia} will only provide PMs for MS stars out to $D\sim 15$ kpc in the halo. Beyond $D\sim 15$ kpc, studies of tangential motion of the stellar halo using \textit{Gaia} PMs will be limited to giants and evolved stars (e.g., \citealt{Bird2018}, \citealt{Lancaster2018}). While giants make excellent tracers due to their bright apparent magnitudes, it is impossible to uniformly select giants from all age and metallicity populations in the halo. Giants are also rare; averaging over large areas of the sky (and thus potential inhomogeneities in the halo) is often required when estimating halo properties with giants.
    
    The HALO7D project seeks to complement the \textit{Gaia} mission by measuring 3D kinematics of distant MW halo MS stars. HALO7D includes both Keck spectroscopy and \textit{HST} PMs for MW halo star candidates in the magnitude range $19<m_{F606W}<24.5$. This dataset provides a deep, densely sampled view of the garden variety stars of the MW halo. In the first HALO7D paper (\citealt{Cunningham2018}; hereafter Paper I), we presented the spectroscopic component of the HALO7D dataset. In this paper, the second in the HALO7D series, we introduce the proper motion component of HALO7D, and use our full 3D kinematic sample to study the halo velocity ellipsoid and anisotropy.
	
	In this work, we seek to use the HALO7D dataset to estimate the parameters of the velocity ellipsoid, and velocity anisotropy, of distant halo MS stars. 	    
    This paper is organized as follows. In Section \ref{sec:data}, we describe the HALO7D dataset and present the HALO7D PM samples. In Section \ref{sec:model}, we describe our methodology for estimating the halo velocity ellipsoid parameters from our observables. In Section \ref{sec:results}, we present our resulting posterior distributions for ellipsoid parameters and velocity anisotropy. In Section \ref{sec:disc}, we compare our results to previous work and other studies. In Section \ref{sec:latte}, we investigate the spatial and radial variation of $\beta$ for two halos from the \textit{Latte} suite of simulations. We conclude in Section \ref{sec:concl}. Details on our computational method for deriving PM uncertainties are given in Appendix \ref{sec:pm_model}; a description of how we tested our ellipsoid parameter model with fake data is given in Appendix \ref{sec:fake_data_ellipsoid}.
	
	\begin{figure*}[h]
\includegraphics[width=\textwidth]{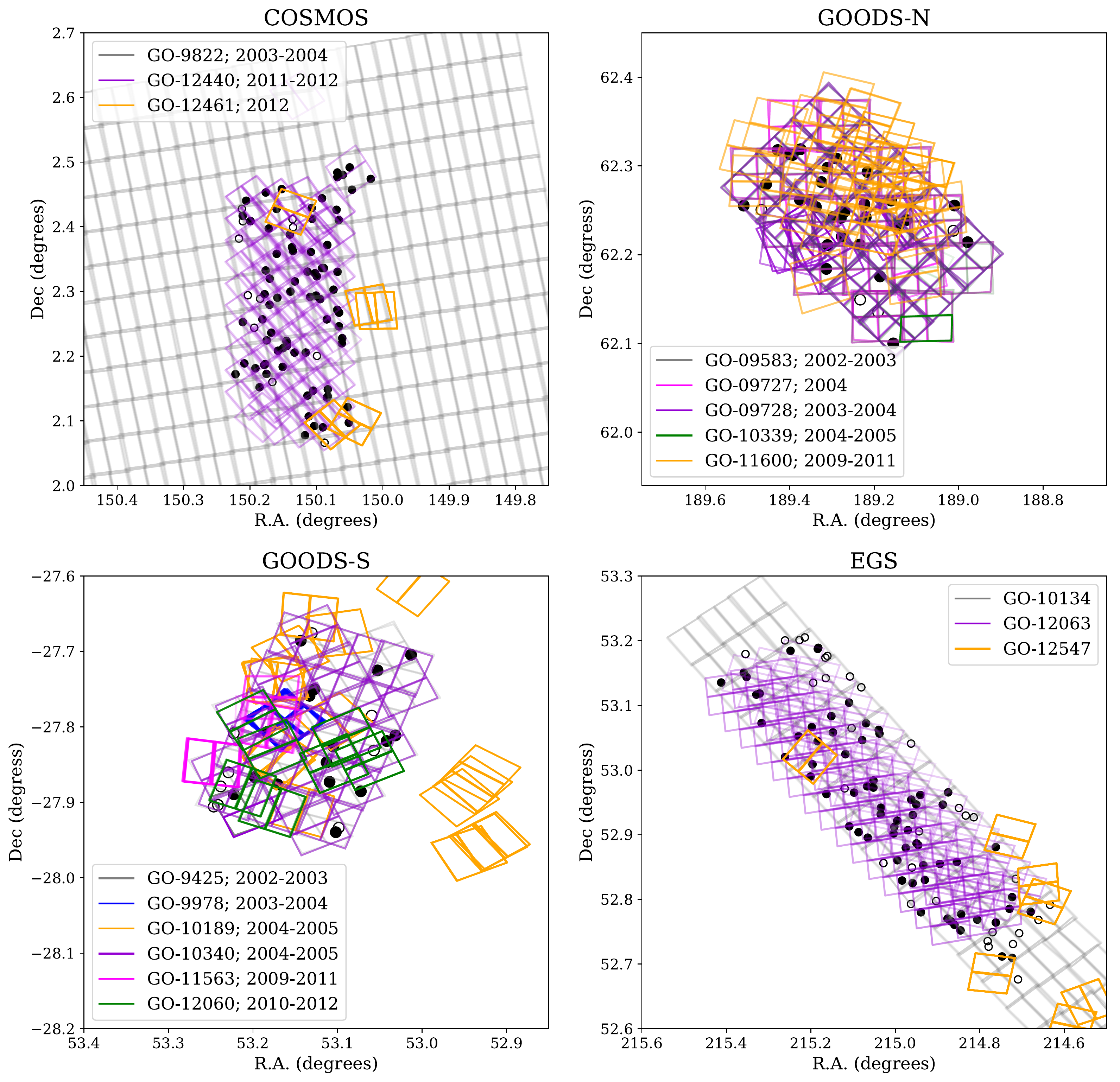}
\caption{The multi-epoch \textit{HST}/ACS footprints of the four HALO7D fields. Different colors indicate the positions of each ACS chip in the different \textit{HST} programs used to measure PMs in this work. HALO7D spectroscopic targets are indicated by black points; filled points indicate targets for which we successfully measured a PM, whereas empty circles indicate targets for which we could not measure a PM.}
\label{fig:footprints}
\end{figure*}
	
	\section{Dataset}
	\label{sec:data} 

HALO7D consists of Keck/DEIMOS spectroscopy and \textit{HST} measured PMs of MW MSTO stars in the EGS, COSMOS, GOODS-N and GOODS-S fields. Coordinates of the HALO7D fields are listed in Table \ref{tab:fields}. We begin this section by summarizing some of the key details on target selection, survey properties, and radial velocity measurements that are discussed in detail in Paper I; the remainder of this section is devoted to a discussion of the proper motion measurements. 

\subsection{Keck/DEIMOS Spectroscopy}

The HALO7D spectroscopic program was described in detail in Paper, I, but we summarize the key details here. 

Candidate halo stars were identified from color-magnitude diagrams. To minimize disk contamination, we selected blue, faint ($19<m_{F606W}<24.5$) objects with star-like morphologies. Stars were observed with Keck II/DEIMOS, configured with the 600ZD grating centered at 7200 \AA, beginning in April 2014 with the final observations taking place in April 2017. We targeted each DEIMOS mask for a minimum of 8 hours of total integration time, and up to 24 hours. 
    
    The radial velocities for these stars were measured using a new Bayesian hierarchical method, called \textsc{Velociraptor}. In order to build up sufficient signal to noise on our targets, stars were observed many times, sometimes over the course of years. Different observations of the same star will have different raw velocities; this is due to the motion of the Earth around the sun (the heliocentric correction) as well as offsets in wavelength solution due to slit miscentering (the A-band correction). We used a Bayesian hierarchical model in order to combine these different observations into a single estimate of the star's velocity. For further details on this technique, we refer the reader to Paper I.
   	
	\subsection{HST Proper Motions}
    
 \begin{figure*}
    \centering
    \includegraphics[width=0.245\textwidth]{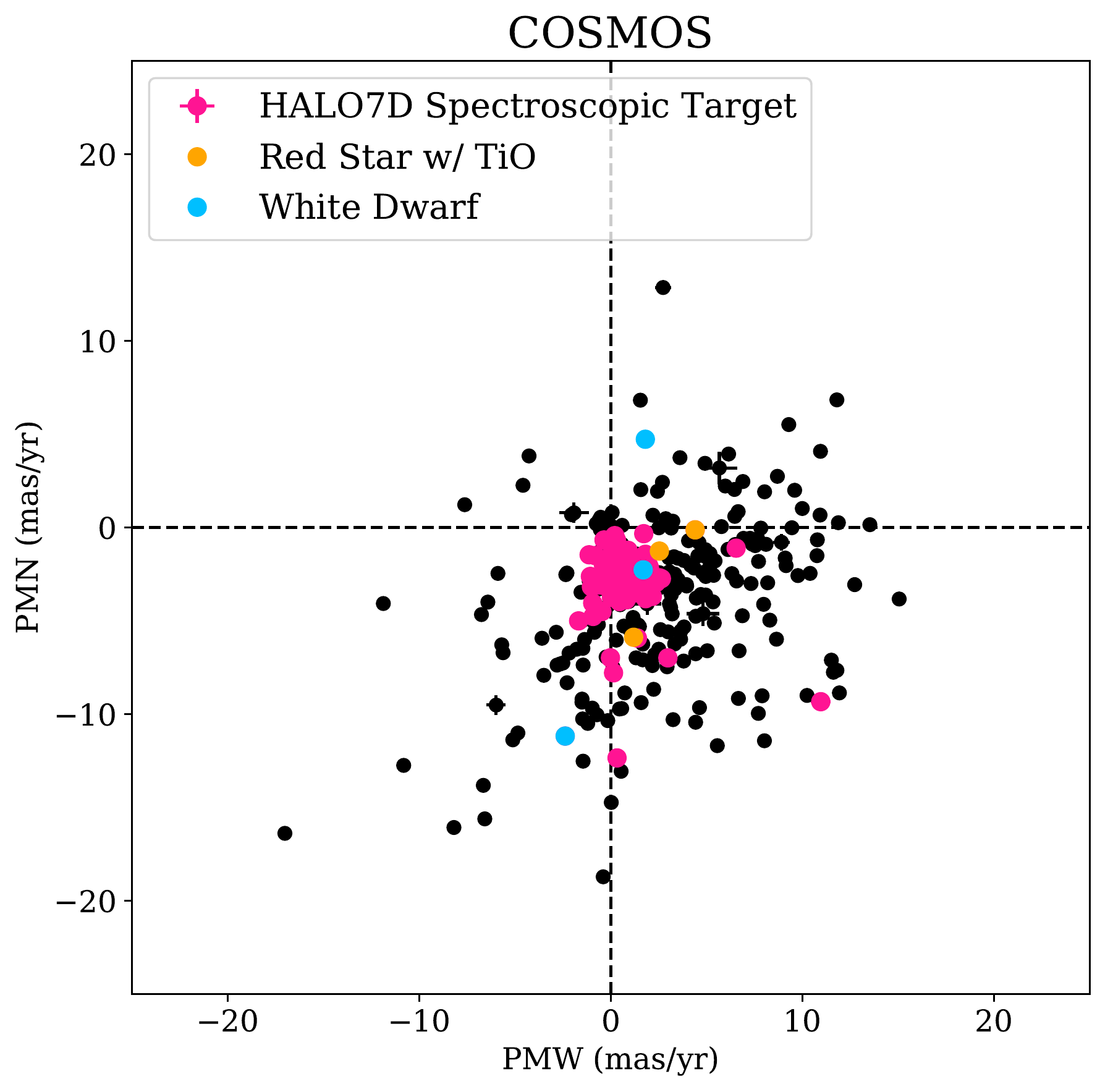}
	\includegraphics[width=0.245\textwidth]{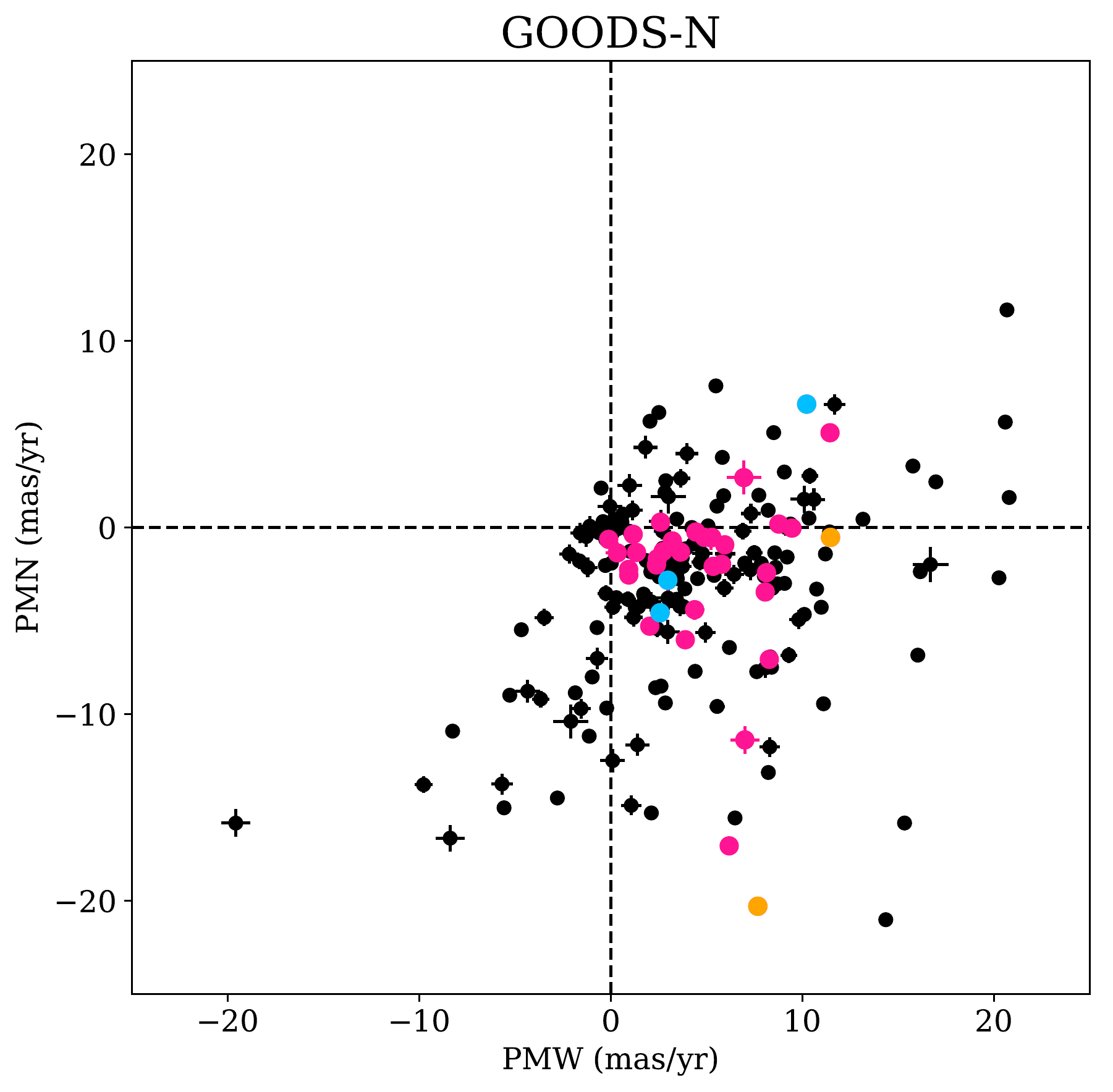}
	\includegraphics[width=0.245\textwidth]{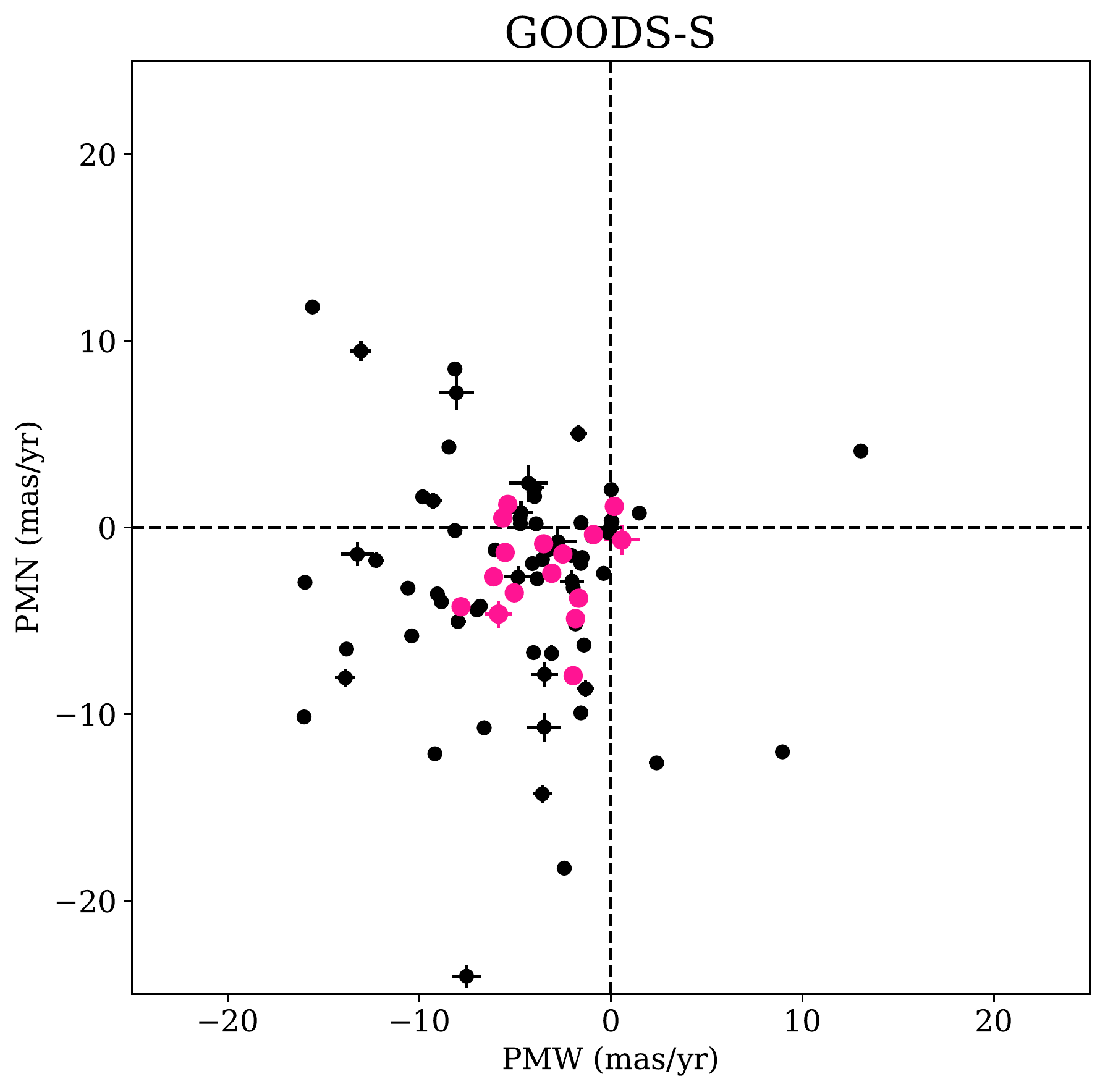}
	\includegraphics[width=0.245\textwidth]{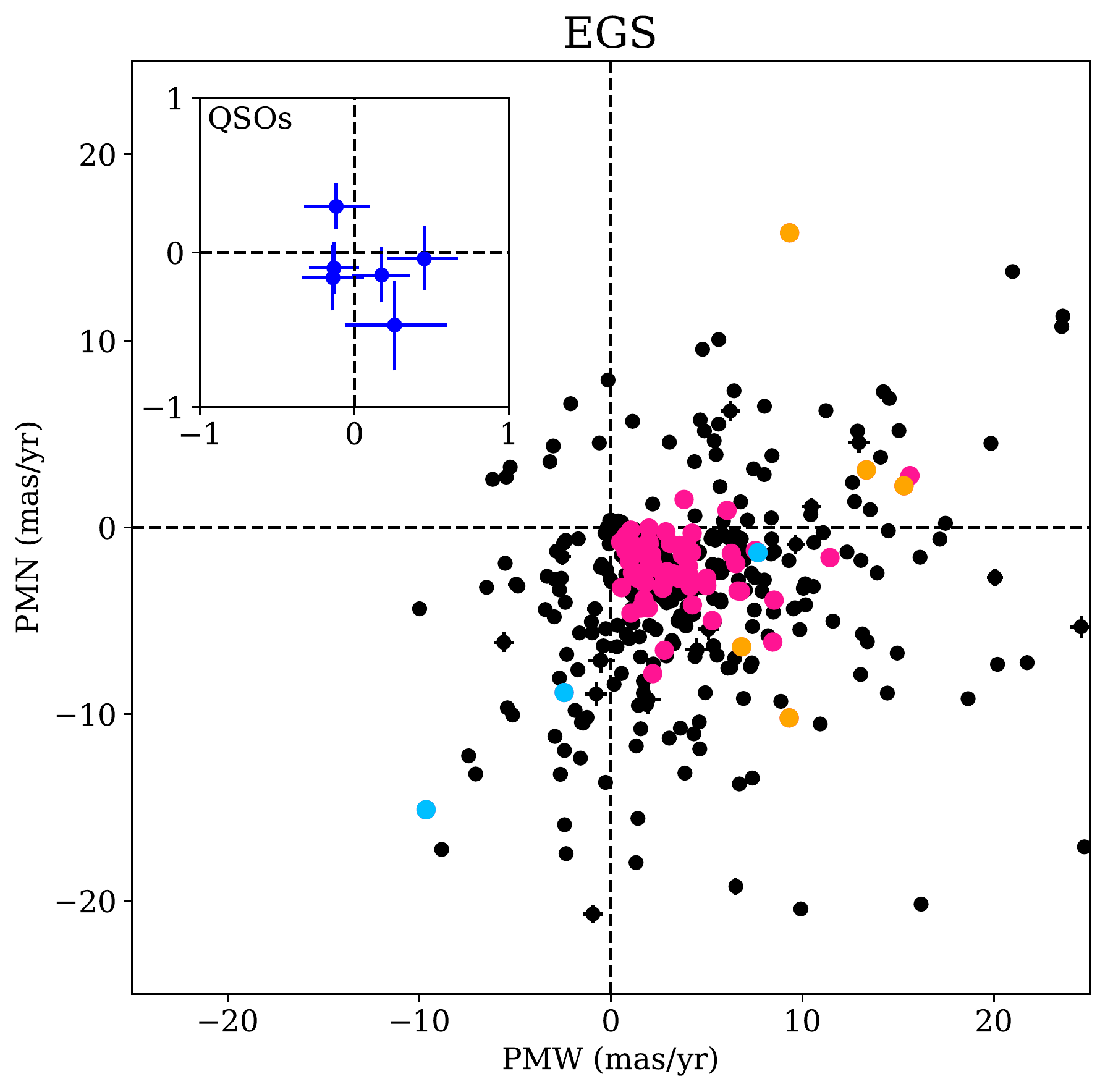}
    
    \includegraphics[width=0.245\textwidth]{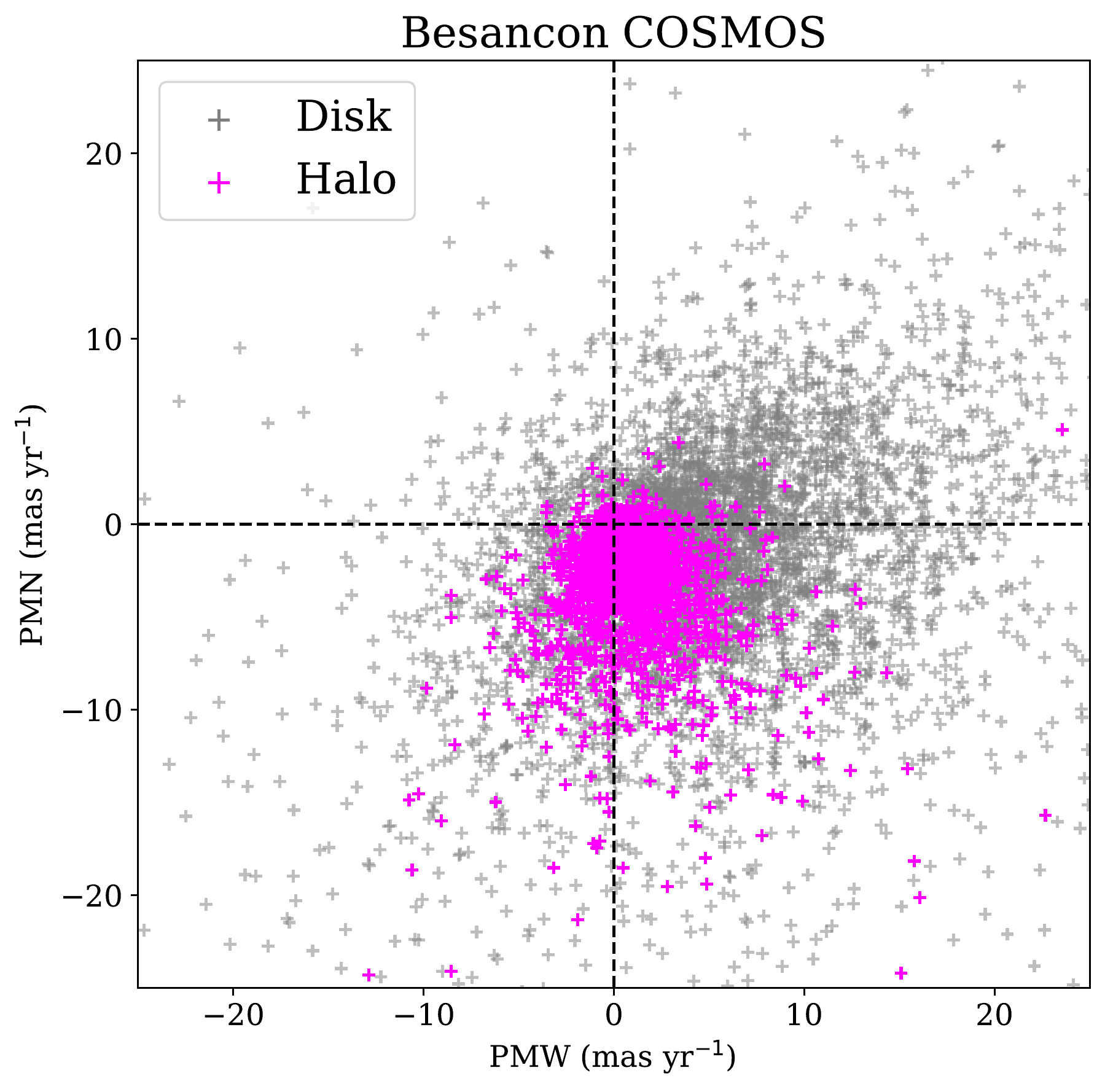}
	\includegraphics[width=0.245\textwidth]{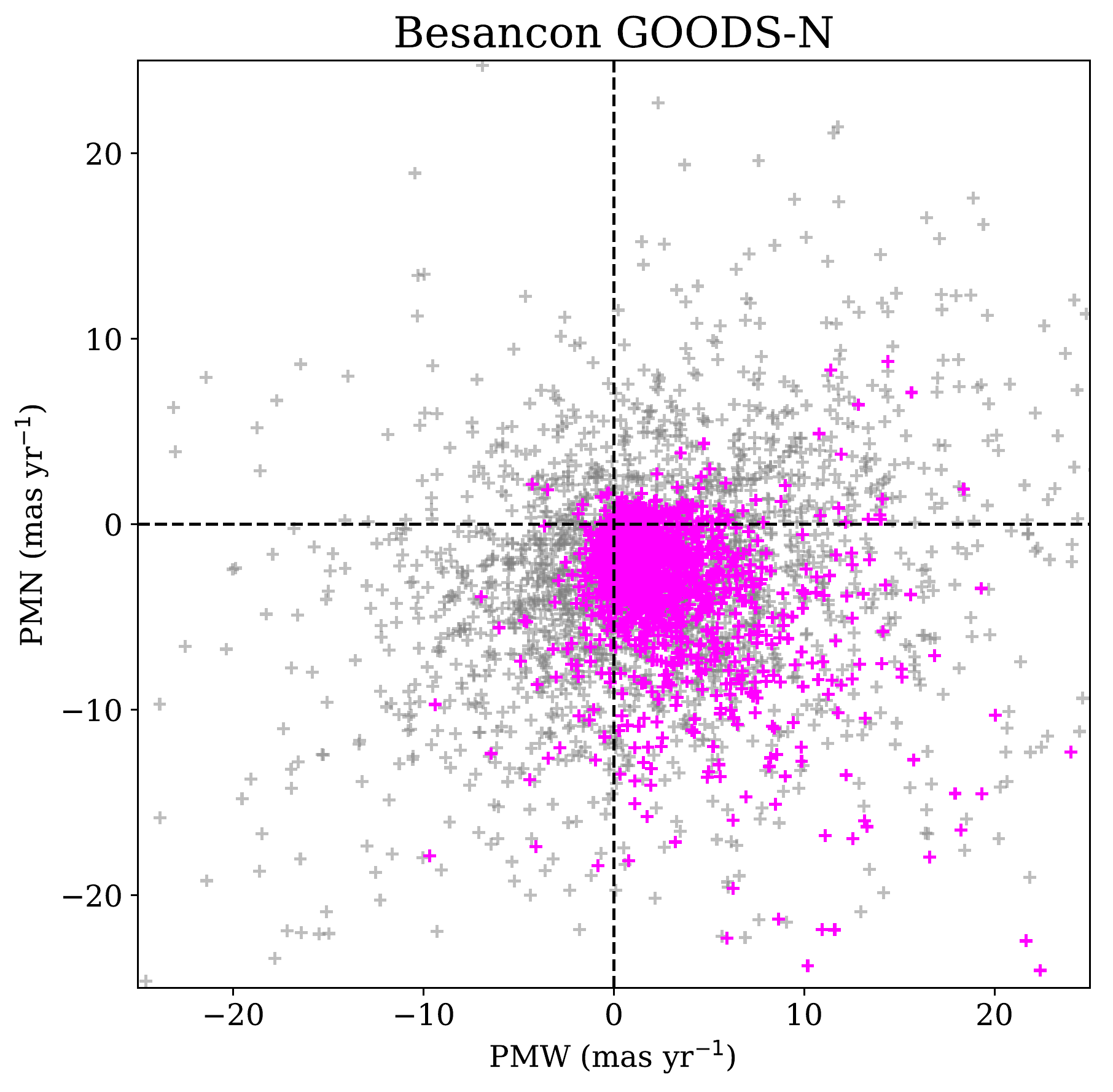}
	\includegraphics[width=0.245\textwidth]{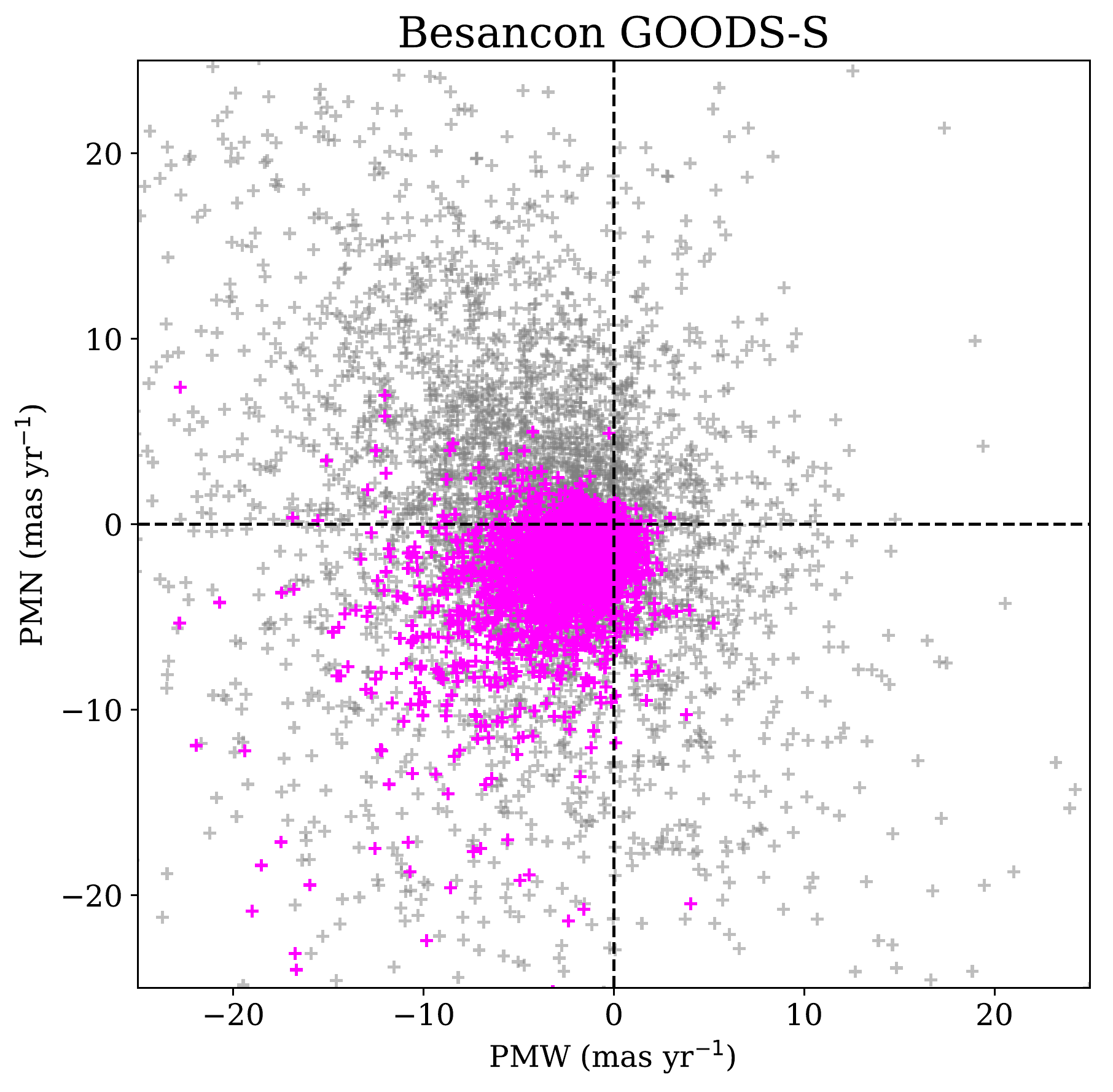}
    \includegraphics[width=0.245\textwidth]{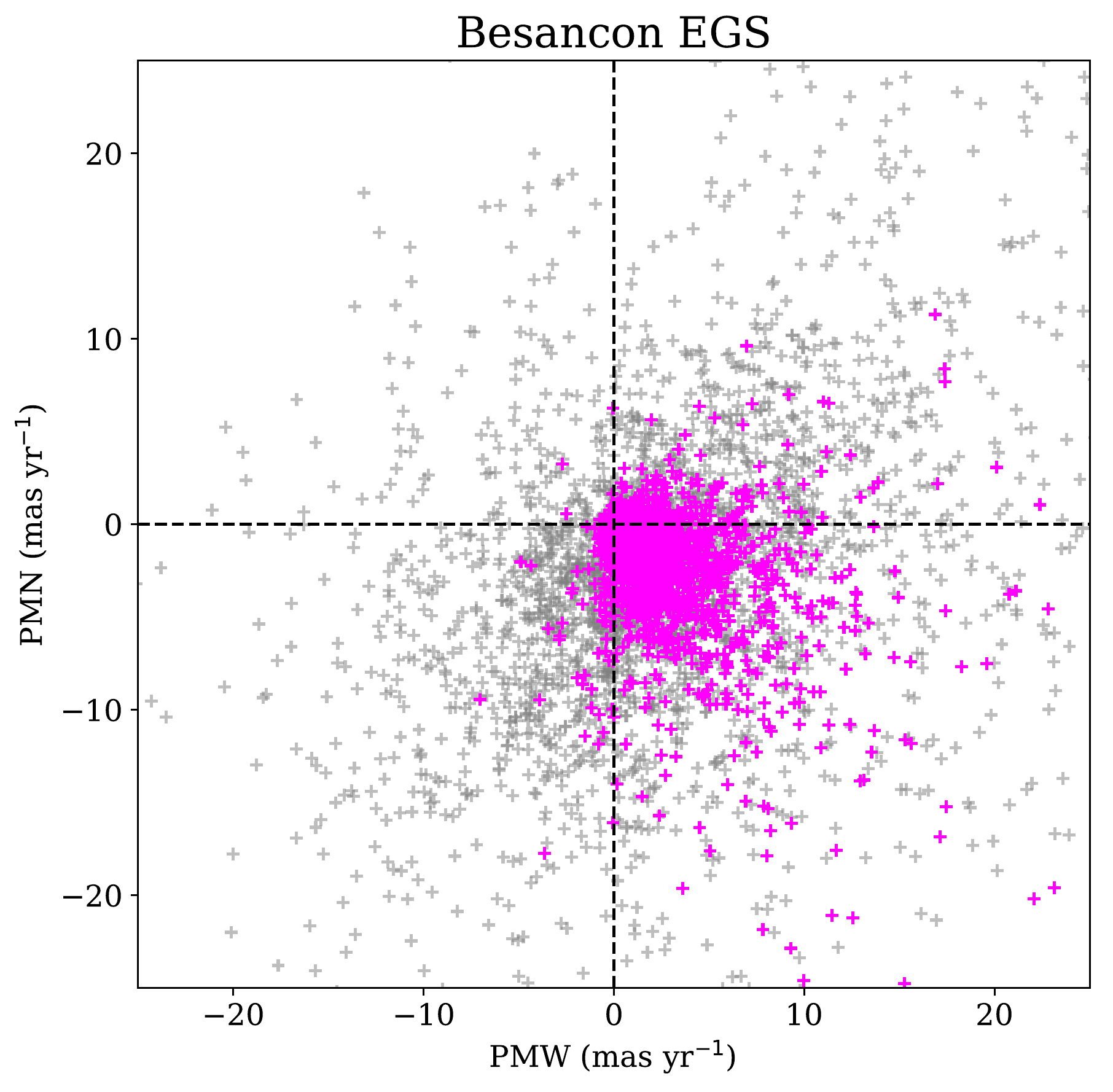}
	
	\caption{Top panels: Proper motion diagrams of the four HALO7D fields. Black points indicate stars that were in the \textit{HST} images that were not HALO7D spectroscopic targets --- primarily MW disk stars. HALO7D spectroscopic targets are shown in pink. Light blue points indicate spectroscopically confirmed WDs, while red MW disk star contaminants are shown in orange. In the EGS PM diagram, the inset shows PMs and errorbars for six spectroscopically confirmed quasars observed in the EGS field. Lower panels: PM diagrams from the Besan\c{c}on Galaxy Model, from 1 square degree fields centered on the coordinates of the HALO7D fields. Grey crosses indicate Besan\c{c}on disk stars, whereas magenta crosses are Besan\c{c}on halo stars.}
	\label{fig:pms}
	\end{figure*}
    
    The measurement methodology in this work builds from existing \textit{HST} PM measurement techniques, relying on the key concept that distant galaxies can be used to construct an absolute stationary reference frame (e.g., \citealt{Mahmud2008}). Sohn et al. (\citeyear{Sohn2012a}, \citeyear{Sohn2013}, \citeyear{Sohn2017}) present detailed descriptions of the state-of-the-art PM measurement techniques used to measure the PMs of Local Group systems with \textit{HST} data. These techniques have been used to measure the PMs of M31 (\citealt{Sohn2012a}), dwarf galaxies Leo I (\citealt{Sohn2013}), Draco and Sculptor (\citealt{Sohn2017}); MW GCs (\citealt{Sohn2018}); and several MW streams (\citealt{Sohn2016}). The PMs of individual MW halo stars measured with \textit{HST} were first published by \cite{Deason2013b}; subsequently, the PMs of individual stars belonging to MW streams were published by \cite{Sohn2015} and \cite{Sohn2016}.
    
    However, the previous Sohn PM studies have typically used only a few \textit{HST} pointings in each study; in that work, they were able to carefully select galaxies by eye that are suitable for use in the reference frame. In order to measure PMs over the full area of the CANDELS fields, we required an approach that could identify ``good" galaxies (with well-measured positions) and ``bad" galaxies (with poorly measured positions) without relying on visual inspection. We therefore built upon existing PM techniques in this work, implementing a Bayesian mixture model that identifies ``good" and ``bad'' galaxies probabilistically and incorporates this uncertainty into the ultimate measurement of the PMs of the stars in the set of images. 

    \subsubsection{Measuring Proper Motions}
    
    In order to measure PMs for the HALO7D targets, we first had to identify the \textit{HST} programs and filters to use for PM measurements. The GOODS, COSMOS, and EGS fields have all been observed multiple times with various setups (detectors + filters). Among them, we selected data that provide astrometric quality sufficient for measuring absolute PMs of individual halo stars. Specifically, data used for our PM measurements meet the following conditions: (1) observations must be obtained with either ACS/WFC or WFC3/UVIS; (2) observations must be in one of the broad-band filters F606W, F775W, F814W, or F850LP; (3) time baseline of the multiple epochs must be at least 2 yr; (4) combined exposure time in the shallower epoch must be at least one orbit long; and (5) individual exposure time must be at least 300 s long. The details of the \textit{HST} programs used for the PM measurements are listed in Table \ref{tab:hst_prgms}; the footprints from these programs are shown in Figure \ref{fig:footprints}. 
    
    Once the \textit{HST} programs and filters were chosen, the \verb|*_flc.fits| images were downloaded from MAST. These images are corrected for imperfect charge transfer efficiency using the algorithms described in \cite{Anderson2010}. The \verb|*_flc.fits| images are processed by a custom-made FORTRAN routine called \verb|*_flc.fits|, which takes a list of RA, Dec positions for objects, identifies them in an flc image, and measures them with a library PSF (see \citealt{Anderson2006}, AK06), determining for each a position, flux, and stellarity index.  The routine then uses the WCS header of each exposure and the distortion solution in AK06 to convert the source positions into an RA-Dec frame.  This routine is run on all the exposures that cover a particular field.

In this analysis, we measure the PMs on a star-by-star basis.  For every target star, the first step is to identify all images that contain the star of interest.  The single-exposure catalogs from the {\tt flt2xym4rd}  output are then fed into another custom-made routine, {\tt xid2mat}, which takes the single-exposure catalogs in pairs and transforms one catalog into the frame of the other, using the galaxy positions as the basis for the transformation.  This transformation makes an initial assessment of which galaxies have consistent positions between the two frames, though the ultimate weighting of the galaxies is done in a Bayesian fashion.
    
    We then specify one image as the reference image: the reference image has the maximum amount of overlap with the other images across epochs containing the star of interest. All overlapping images are mapped onto the reference image frame with \verb|xid2mat| using a six-parameter linear transformation:
	
		\begin{equation}
			\begin{pmatrix}
				A && B && x_t \\
				C && D && y_t \\
				0 && 0 && 1   \\
			\end{pmatrix}
			\quad
			\begin{pmatrix}
				&u& \\
				&v& \\
				&1& \\
			\end{pmatrix}
			=
			\begin{pmatrix}
				u_{ref} \\
				v_{ref} \\
				1 \\
			\end{pmatrix},		
            \label{eqn:trans}
		\end{equation}
        where $(u,v)$ are the vectors of distortion corrected positions of objects in one image and $(u_{ref},v_{ref})$ are the vectors of positions in the reference image. The parameters $x_t,y_t$ represent any linear translation offset between the two images, while parameters $A,B,C,D$ incorporate scale, rotation, and off-axis linear camera distortion terms. The positions of stars are used to match frames within an epoch, and the positions of ``good" galaxies are used to match images across epochs. For more details on why these transformations are required in comparing \textit{HST} images, please see section 3.6.4 in \cite{Anderson2010b}. 

When the images across epochs have been matched via the linear transformation, the change in the positions of the stars across epochs provides an initial estimate of their PMs. In order to get full posterior probability distributions for the PMs, and incorporate all sources of uncertainty (such as uncertainty in star and galaxy positions, as well as which galaxies should be including in the stationary reference frame), we use a Bayesian mixture modeling approach. We leave as free parameters the positions of all stars and galaxies, the image transformation parameters, and the proper motions of all stars. We model the galaxies in our reference frame as being a mixture of ``good" and ``bad" galaxies (with poorly measured positions). Within an epoch, we use the positions of stars to precisely align the images. 

Table \ref{tab:fields} lists the resulting median PM errors in each of the HALO7D fields. Our PM errors are not a function of the magnitudes of our stars, but rather our ability to define the stationary reference frame for a given target. This is determined by how many images there are containing a given star, how much these images overlap across epochs, and how many ``good" galaxies there are in the images. For a full description of the Bayesian model for this problem, as well as the details of the Gibbs sampling algorithm we used to sample from the full posterior, we refer the reader to Appendix \ref{sec:pm_model}. 

\begin{table*}
	\begin{center}
	\begin{tabular}{c  c  c  c  c c c }
		Field & \textit{l} (deg) & \textit{b} (deg) & $v_{l, \odot}$ (km s$^{-1})$ & $v_{b, \odot}$ (km s$^{-1}$ )& Median PM Error (mas yr$^{-1}$) \\
		\hline \hline
        COSMOS & $236.8$ & $42.1$ &  $-126.0$& $148.2$ & $0.16$\\
        GOODS-N &$125.9$ & $54.8$ & $-153.8$ & $-154.2$& $0.12$\\
        GOODS-S & $223.6$ & $-54.4$ & $-171.5$ & $-140.8$ &$0.28$\\
        EGS & $96.4$ & $60.4$ &$ -38.7 $&$ -209.0$& $0.18$\\
	\end{tabular}
	\end{center}
	\caption{Galactic coordinates, projection of the Sun's velocity in Galactic coordinates, and the median PM error (in Galactic coordinates) for the four HALO7D fields. Quoted median PM errors are the errors in a single component (e.g., $\mu_l \cos(b)$ or $\mu_b$; we find both components of PM have the same median errorbars, to within 0.005 mas yr$^{-1}$, within a given field).}
	\label{tab:fields}

\end{table*}

\begin{figure*}
\centering
	\includegraphics[width=0.8\textwidth]{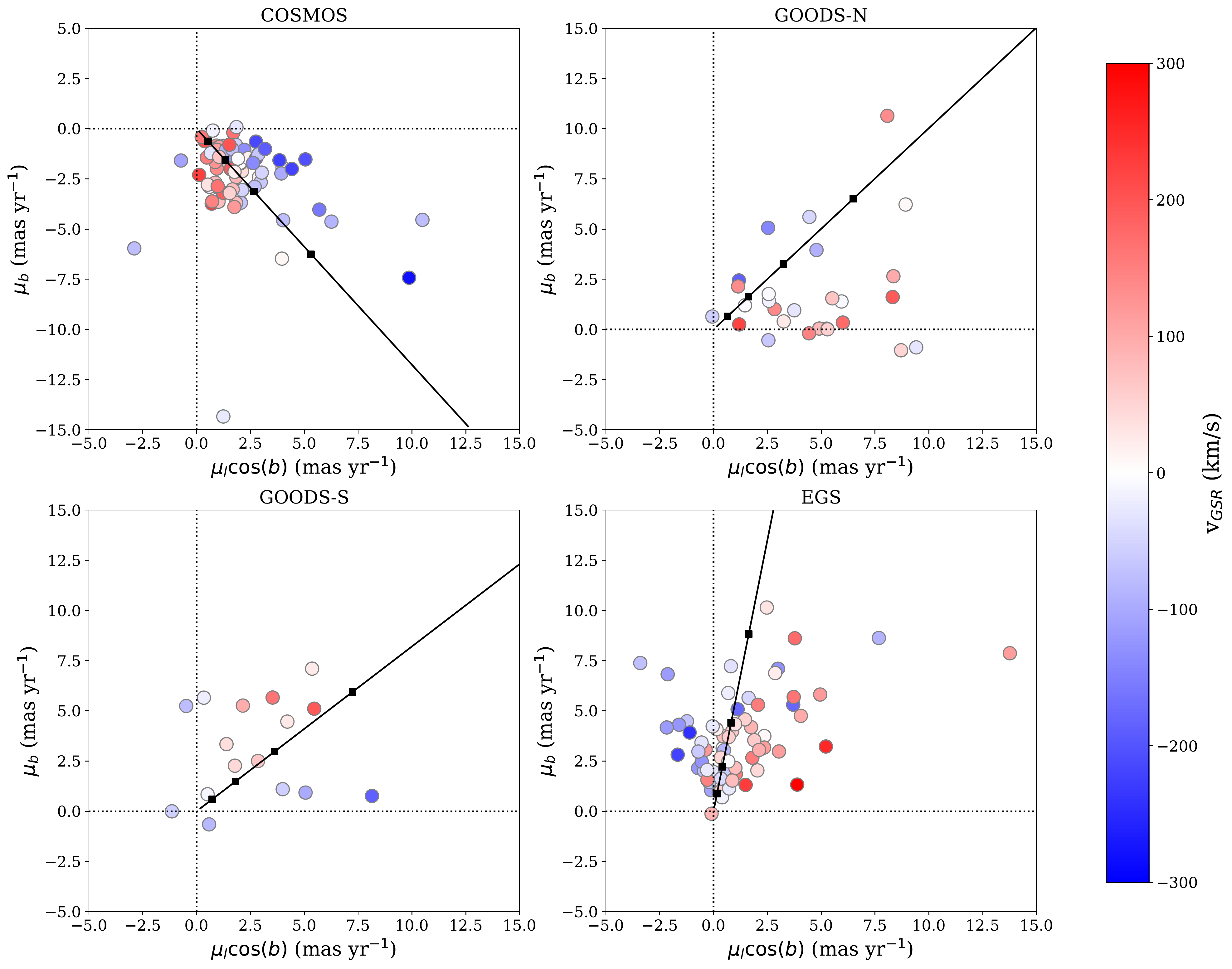}
	\caption{Proper motion diagrams of the four HALO7D fields, in Galactic coordinates, color coded by LOS velocity in the Galactocentric frame. Solid black lines indicate the solar reflex motion along each sightline; the squares indicate the implied mean PM along each line of sight for $D=5,10,20,50$ kpc (with mean PM at larger distances approaching ($\mu_l\cos(b),\mu_b)=(0,0)$. Dotted lines indicate $\mu_l\cos(b)=0~\mathrm{mas~yr}^{-1}, \mu_b=0~\mathrm{mas~yr}^{-1}$. }
	\label{fig:3d}
	\end{figure*}

    \subsubsection{Proper Motion Diagrams}
	
	Figure \ref{fig:footprints} shows positions of the HALO7D spectroscopic sample as black points; filled circles indicate targets for which we successfully measured PMs, and open circles are stars for which we couldn't measure a PM. As can be seen in Figure \ref{fig:footprints}, the \textit{HST} pointings from different epochs are not well aligned; this is because we are using archival data for \textit{HST} programs that were not designed with astrometry in mind. As a result, some of the HALO7D targets only have one epoch of \textit{HST} imaging. This usually arises when the target is on the edge of the field, or if the target falls in the ACS chip gap in one of the epochs. 
    
    PM diagrams for the four HALO7D fields are shown in the top panels of Figure \ref{fig:pms}. PMs are plotted in $\mathrm{PMW}=-\mu_{\alpha} \cos(\delta)$, $\mathrm{PMN}=\mu_{\delta}$. The PMs of HALO7D halo star candidates are shown in pink. Our PM method returns PMs and uncertainties for all point-like objects in the specified reference image that have multi-epoch coverage; PMs for objects that were not HALO7D spectroscopic targets are shown as black points. Most of these points are MW disk stars, though a few will be point-like distant galaxies. As explained in Paper I, our spectroscopically confirmed disk contaminants are white dwarfs (WDs) and red stars with titanium oxide absorption features. These disk contaminants are shown as light blue and orange points, respectively. The black, light blue and orange points occupy a larger area of PM space than the pink points; because they are mostly disk members, they are at closer distances than the HALO7D halo star candidates and thus have higher proper motions. The PM diagrams as predicted by the Besan\c{c}on Galaxy Model (\citealt{Robin2003}), for one square degree fields centered on our field coordinates, are shown in the lower panels of Figure \ref{fig:pms} for reference. 
	
	In addition, in the EGS field, we had six spectroscopically confirmed quasars for which we could also measure PMs. These PMs are shown in the inset of the upper lefthand panel of Figure \ref{fig:pms}; reassuringly, all quasar PMs are consistent with 0 mas yr$^{-1}$.
	
	\begin{table*}
	\begin{center}
	\begin{tabular}{c|  c  c  c  c}
    \hline
	 Field & Program & P.I. & Filter & Dates \\
     \hline
     \hline
    COSMOS & GO-9822 &Scoville  & F814W & 12/2003--05/2004\\
    & GO-12440 & Faber & F814W & 12/2011--02/2012\\
    & GO-12461 & Riess & F814W &02/2012 -- 04/2012 \\
	\hline
    GOODS-N & GO-9583& Giavalisco& F775W & 11/2002--05/2003 \\
    &GO-9727 & Perlmutter & F775W & 04/2004--08/2004\\
    &GO-9728 & Riess & F775W & 06/2003--09/2004\\
    &GO-10339 & Riess & F775W & 10/2004--04/2005\\
    &GO-11600 & Weiner & F775W & 9/2009--04/2011\\
    \hline
    GOODS-S & GO-9425 & Giavalisco & F606W, F850LP & 07/2002--02/2003\\
    & GO-9978 & Beckwith& F606W, F850LP & 09/2003 -- 01/2004\\
    & GO-10189 & Riess &F606W, F850LP & 09/2004 -- 08/2005\\
    & GO-10340 & Riess&F606W, F850LP & 07/2004 -- 09/2005\\
    & GO-11563 & Illingworth & F606W, F850LP &08/2009 -- 02/2011 \\
    & GO-12060/1/2 & Faber & F606W, F850LP &08/2010 -- 02/2012 \\
    \hline
     EGS & GO-10134 & Davis & F814W & 06/2004 -- 03/2005 \\
     & GO-12063 & Faber & F814W & 04/2011 -- 05/2013 \\
     & GO-12547 & Cooper & F814W & 10/2011 -- 02/2013 \\
    \hline
	\end{tabular}
	  \caption{Summary of the \textit{HST} programs used for the PM measurements in this paper.}
	\label{tab:hst_prgms}
	\end{center}
	\end{table*}

	Our 3D kinematic sample is summarized in Figure \ref{fig:3d}; PM components in $(l,b)$ are plotted against each other for the four HALO7D fields, color coded by LOS velocity as measured in Paper I. In EGS, we see an interesting covariance between $\mu_l$ and $v_{LOS}$; there appears to be a trend of increasing LOS velocity with increasing $\mu_l$. However, we note that this is not a signature of rotation. EGS is located at a Galactic longitude $l=96$ degrees; along this line of sight, $v_l \approx -V_X$, and $V_Y=v_{LOS} \cos(b)-v_b\sin(b)$. Given that $v_{\phi}=\frac{x}{R_p}V_Y-\frac{y}{R_p}V_X$, a covariance between $V_X, V_Y$ arises naturally if we assume a Gaussian velocity distribution for $v_{\phi}$. As we'll see in Section \ref{sec:results}, the fact that $v_{LOS}$ and $\mu_{l}$ increase together is consistent with zero net rotation along this line of sight.

	\section{Modeling The Halo Velocity Ellipsoid}
	\label{sec:model}
	
	We use our 3D kinematic sample to estimate the parameters of the halo velocity ellipsoid in spherical coordinates. In this work, we use only objects for which we have both a successful PM measurement and a successful LOS velocity measurement; we leave the analysis of stars with PM measurements but without LOS velocities to future work. Our method is very similar to the ones used in \cite{Cunningham2016} and \cite{Deason2013b}, though in this work we have used notation and language consistent with a Bayesian construction of the problem. 
	
	For each star $i$ located in field $k$, we have data $y_i=\{v_{LOS},\mu_l, \mu_b\}$, with associated explanatory variables $x_i=\{m_{F606W,i}, m_{F814W,i}, l_k, b_k\}$. We model our sample as being drawn from a mixture of two distributions: the disk distribution (with fixed parameters), and the halo distribution. 
    
    The free parameters in our model are the absolute magnitudes (and, by extension, the distances) to each star $\textbf{M}=\{ M_{F814W,1} ,..., M_{F814W,N} \}$ (we denote the corresponding distances $\textbf{D}=\{ D_1 ,..., D_N \}$); the fraction of disk contamination along a given line-of-sight $\textbf{f}=\{f_{\mathrm{Disk},1}, ..., f_{\mathrm{Disk}, k}\}$; and the halo velocity ellipsoid parameters $\theta_{\rm Halo}=\{\langle v_{\phi} \rangle, \sigma_r, \sigma_{\phi}, \sigma_{\theta}\}$. 
	
	\subsection{Disk Model}
    \label{sec:disk}
	
	For the disk model, we work in cylindrical coordinates $(R_p, \phi, z)$. We assume exponential density profiles in both $R_p$ and $z$, with a disk scale length of $h_{R}=3$ kpc and a disk scale height of $h_z=1$ kpc.

	For the disk velocity distributions, we assume distributions in $R_p$ and $z$ that are Gaussian with zero net motion, and have dispersions of $\sigma_{R_P}=45$ km s$^{-1}$ and $\sigma_z = 20$ km s$^{-1}$. For the tangential component, we assume that the rotational velocities are described by a skewed normal distribution with mean $\langle v_T \rangle = 242$ km s$^{-1}$, scale parameter $46.2$ km s$^{-1}$, and shape parameter of $-2$. These parameters are derived based on the predicted marginalized velocity distributions from \verb+galpy+\footnote{http://github.com/jobovy/galpy} (\citealt{Bovy2015}), using the quasi-isothermal distribution function discussed in \cite{Binney2010} and \cite{Binney2011} and the \verb+MWPotential2014+ (see \citealt{Bovy2015} for details). While they are not free parameters in our model, for simplicity in notation, we denote the disk DF parameters as $\theta_{\rm Disk}$.
	
	While this disk model is quite simple, we find that this model is effective at identifying stars in our sample that are disk-like (see Section \ref{sec:disk_results}). These stars are more likely to have higher proper motions, brighter apparent magnitudes, redder colors, and heliocentric LOS velocities closer to 0 km s$^{-1}$. 
	
	\subsection{Halo Model}
	
	For the halo distribution, we work in spherical coordinates. We assume the broken halo density profile derived in \cite{Deason2011b}, with break radius $r_b=27$ kpc and slopes $\alpha_{\rm{in}}=2.3$, and $\alpha_{\rm{out}}=4.6$. The probability that a star has a distance $D_i$ given the density profile is given by: 
	
	\begin{equation}
		p(D_i|\rho,l,b) \propto \rho(r_q(D_i,l,b)) \times D_i^2 ,
	\end{equation}
where the factor of $D_i^2$ arises from the spatial volume element in spherical coordinates.
	
	We assume independent Gaussian velocity distributions for the three spherical components of motion, and assume $\langle v_r \rangle = \langle v_{\theta} \rangle=0$ km s$^{-1}$. We define our vector of halo ellipsoid parameters to be $\theta_{\rm Halo}=\{ \langle v_{\phi} \rangle, \sigma_r, \sigma_{\phi}, \sigma_{\theta} \}$. We denote joint velocity PDF:
	
	\begin{equation}
    \begin{split}
		F_{v,\mathrm{Halo}}(v_{r,i},v_{\phi,i},v_{\theta,i}) = \mathrm{N}(v_{r,i}|0, \sigma_r^2+\sigma_{v_r,i}^2) \times \\
        \mathrm{N}(v_{\phi,i}|\langle v_{\phi} \rangle, \sigma_{\phi}^2+\sigma_{v_{\phi},i}^2) \times 
         \mathrm{N}(v_{\theta,i}|0, \sigma_{\theta}^2+\sigma_{v_{\theta},i}^2),
        \end{split}
	\end{equation}
where $v_{r,i},v_{\phi,i},v_{\theta,i}$ are the Galactocentric velocities corresponding to data $y_i$ and distance $D_i$. The corresponding uncertainties on these velocities are denoted by $\sigma_{v_{r},i}^2, \sigma_{v_{\phi},i}^2,\sigma_{v_{\theta},i}^2$. Proper motions in Galactic coordinates are converted to physical velocities using the fact that tangential velocity is proportional to distance: $v_T=4.74047\mu D$, where $\mu$ is the proper motion in mas yr$^{-1}$ and $D$ is in kpc. Tangential velocities are converted to the Galactocentric frame by correcting for the projection of the Sun's velocity along a given line-of-sight. We convert $(v_{LOS}, v_l, v_b)$ to spherical coordinates $(v_r,v_{\phi},v_{\theta})$ by assuming a circular speed of 240 km s$^{-1}$ at the position of the Sun ($R_0=8.5$ kpc), with solar peculiar motion $(U,V,W)=(11.1, 12.24, 7.25)$ km s$^{-1}$ (\citealt{Schonrich2010}). 

	We note that in order to evaluate the probability of $\theta_{\rm Halo}$ given our observables, we need to consider the Jacobian matrix from the coordinate transformation from the observed frame to the Galactocentric frame:
	
	\begin{equation}
		p(y_i |D_i,\theta_{\rm Halo}) \propto F_{v,\mathrm{Halo}} \times D_i^2 \cos{b},
	\end{equation}
where the factor of $D_i^2 \cos{b}$ arises due to the change in variables. 
    
    \subsection{Absolute Magnitudes}
	
	Finally, as in \cite{Deason2013b}, \cite{Cunningham2016}, and in Paper I, we additionally constrain the absolute magnitude to a given star $M_{F814W,i}$ using information on its $m_{F606W}-m_{F814W}$ color.
	
	We weight \cite{Vandenberg2006} isochrones in the \textit{HST} filters according to the approximate age and metallicity distributions of the MW halo. We then generate a KDE to get the probability distribution function $G(M_{F814W}|m_{F606W,i}, m_{F814W,i})$. 
	
	\subsection{Full Posterior}
	
	We now summarize how we sample from our full posterior distribution, for our parameters $\theta_{\rm Halo}, \textbf{M}, \textbf{f}$ given observables $\textbf{y}, \textbf{x}$. We can write down the likelihood under this model for a star with data $y_i$, explanatory variables $x_i=\{m_{F606W,i}, m_{F814W,i}, l_k, b_k\}$, given our model parameters:

    \begin{equation}
		\begin{split}
		p(y_i|\theta_{\rm Halo}, M_{F814W,i}, f_{\mathrm{Disk},k}, x_i) =\\
        p(M_{F814W,i}|m_{F606W,i}, m_{F814W,i}) \times D_i\\
        \times \big[ f_{\mathrm{Disk},k}\times p(y_i|\theta_{\rm Disk}, D_i) p(D_i|\rho_{\rm Disk}, l_k, b_k) + \\
		(1-f_{\mathrm{Disk},k}) \times p(y_i | \theta_{\rm Halo}, D_i) p(D_i|\rho_{\rm Halo}, l_k, b_k) \big] ,\\
		\end{split}
		\label{eqn:mix_like}
	\end{equation}
	
The extra factor of $D_i$ arises due to the change of variables from absolute magnitude to distance: $M_{F814W} \propto \log(D)$. The full likelihood, using stars from $k=1,...,K$ fields containing $N_{*,k}$ stars, is given by the product of the likelihoods of each individual data point:
    
    	\begin{equation}
		\begin{split}
		p(\textbf{y}|\theta_{\rm Halo}, \textbf{M}, \textbf{f}, \textbf{x}) = \prod_{k=0}^K \prod_{i=0}^{N_{*,k}} p(y_i|\theta_{\rm Halo}, M_{F814W,i}, f_{\mathrm{Disk},k}, x_i).
		\end{split}
		\label{eqn:full_mix_like}
	\end{equation}
		
	Likelihood in hand, we can write down the posterior distribution for our model parameters using Bayes Theorem: 
		
	\begin{equation}
	p(\theta_{\rm Halo}, \textbf{M}, \textbf{f}|\textbf{y}) \propto p(\textbf{y}, \textbf{M}|\theta_{\rm Halo}, \textbf{f}) \times p(\theta_{\rm Halo}, \textbf{f}) ,
	\label{eqn:post}
	\end{equation}
where $p(\theta_{\rm Halo}, \textbf{M}, \textbf{f})$ is the prior distribution on model parameters. We assume standard reference priors on $\theta_{\rm Halo}$ (i.e., Jeffreys priors: $p(\langle v_{\phi} \rangle) \propto$ const and $p(\sigma) \propto 1/\sigma$ for all dispersions). We assume uniform priors on the $f_{\rm Disk}$ parameters ($p(f_{\mathrm{Disk},k})=1, f_{\mathrm{Disk},k} 
~\ \epsilon ~\ [0,1]$). 
	
	In order to sample for our model posterior parameters, we compute Equation \ref{eqn:post} over a grid in absolute magnitude for every star. We then use \verb+emcee+ (\citealt{Foreman-Mackey2013}) to sample from our full posterior, marginalizing over the absolute magnitude of every star in each step of the chain. We test this modeling procedure on fake data; for details on how we generated fake data and tested our model, we refer the reader to Appendix \ref{sec:fake_data_ellipsoid}.

\begin{figure*}
		\centering
		\includegraphics[width=0.8\textwidth]{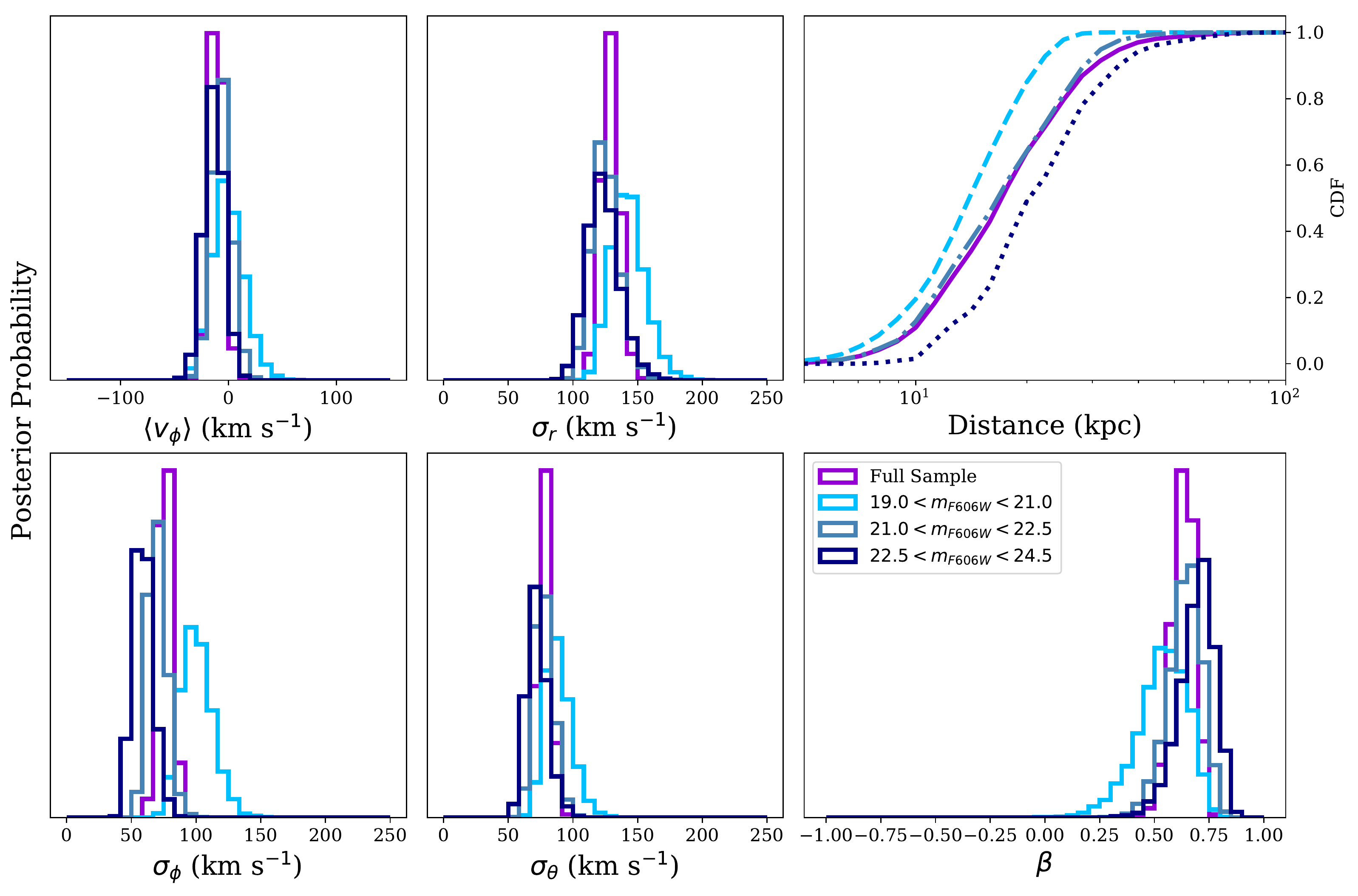}
		\caption{Summary of posterior results for spherically averaged samples. Left-hand panels: histograms of marginalized posterior samples for the four parameters of the halo velocity ellipsoid. Each of the estimates shown combines targets from all four survey fields. The estimates using the full HALO7D sample are shown in purple, while the blue histograms show the resulting estimates from three apparent magnitude bins. Upper righthand panel: CDFs of the distances of the full sample and the three apparent magnitude bins. Lower righthand panel: posterior distributions for $\beta$.}
		\label{fig:sphere_sum}
	\end{figure*}
    
        \begin{figure}
		\centering
		\includegraphics[width=0.5\textwidth]{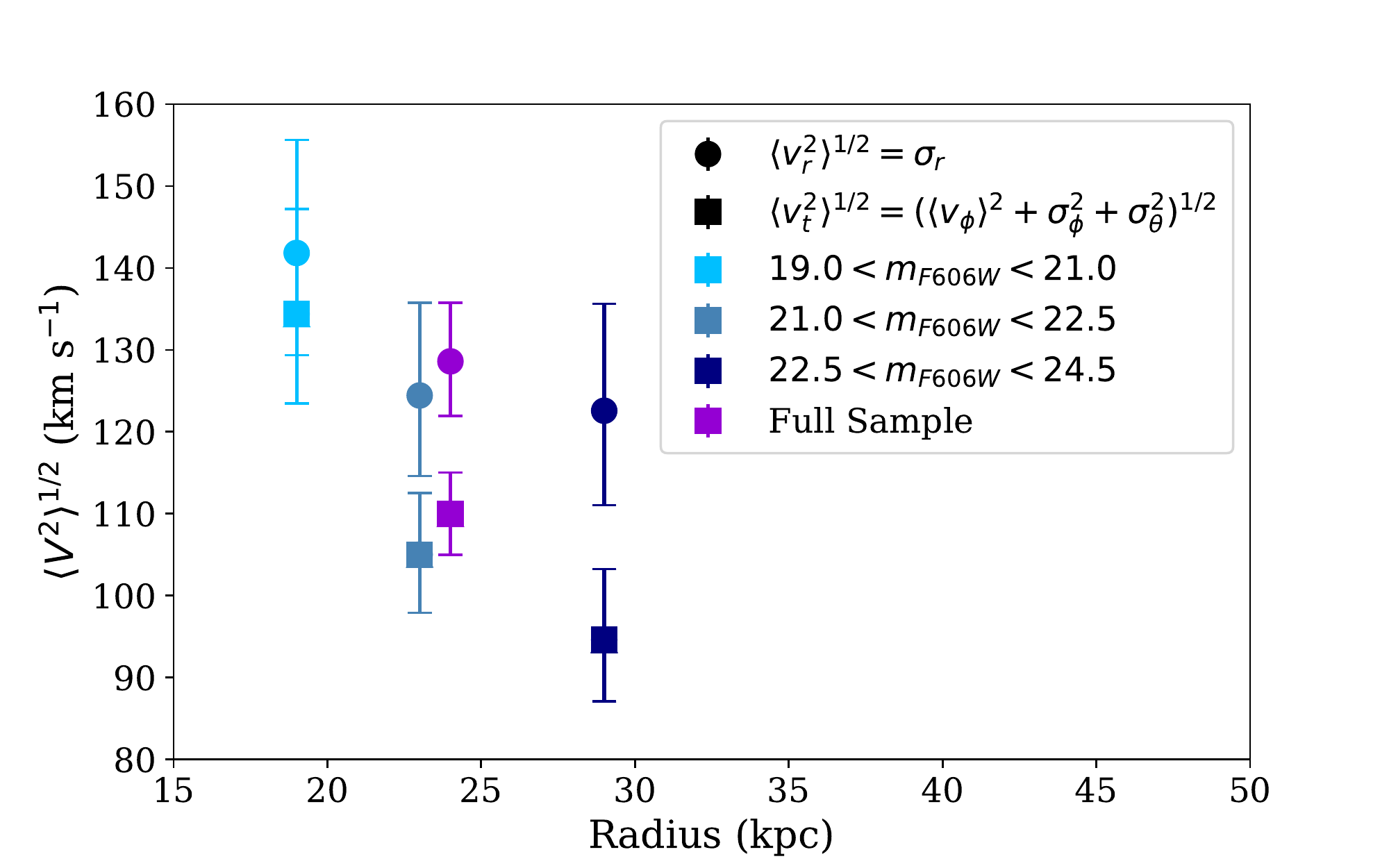}
		\caption{Square roots of the second moments of the radial (circular points) and tangential (square points) velocity distributions, as a function of mean Galactocentric radius. Different colors indicate the results from the analysis of the full sample (purple) as well as the three apparent magnitude bins (same colors as in Figure \ref{fig:sphere_sum}).}
		\label{fig:params_v_d}
	\end{figure}

	\section{Results}
	\label{sec:results}

	In this section, we present posterior distributions for the halo velocity ellipsoid parameters. We first present the results using the full HALO7D sample, and then split our sample into three apparent magnitude bins. Finally, we consider the samples from each field separately.

\begin{table*}
\begin{center}
\begin{tabular}{c|  c  c  c  c  c  c c c c}
 & $\langle v_{\phi} \rangle$ (km s$^{-1}$)& $\sigma_{\phi}$ (km s$^{-1}$)& $\sigma_{\theta}$ (km s$^{-1}$)& $\sigma_{r}$ (km s$^{-1}$)& $\langle D \rangle$ (kpc) & $\langle r \rangle$ (kpc) & $\beta$ & $N_{Stars}$ \\
\hline
Full Sample & $-11 \pm 6$ & $76 \pm 5 $ &$78 \pm 5$ &$129 \pm 7$ &$20$ &$24$ & $0.63 \pm 0.05$& 188 \\
\hline
$19.0 < m_{F606W} < 21.0$ & $-1^{+15}_{-14} $ & $100^{+13}_{-11}$ &$87^{+11}_{-9}$ &$142^{+14}_{-12}$ & $16$& $19$& $0.54^{+0.11}_{-0.12}$& 67\\
$21.0 < m_{F606W} < 22.5$& $-6 \pm 9 $ & $70^{+8}_{-7}$ &$77^{+8}_{-7}$ &$124^{+11}_{-10}$ &$19$ & $23$ & $0.64^{+0.07}_{-0.09}$&  71 \\
$22.5 < m_{F606W} < 24.5$ &$-13^{+9}_{-10} $ & $59^{+8}_{-7}$ &$72^{+9}_{-7}$ &$123^{+13}_{-12}$ & $24$& $29$& $0.70^{+0.09}_{-0.07}$& 50 \\
 \hline
 COSMOS & $-15^{+10}_{-9} $ & $73^{+8}_{-7} $ &$60^{+7}_{-6}$ &$121^{+11}_{-9}$ & $20$& $25$& $0.67^{+0.06}_{-0.08}$& 77 \\
 GOODS-N & $10^{+22}_{-21} $ & $100^{+19}_{-16} $ &$122^{+19}_{-16}$ &$142^{+22}_{-17}$ & $23$& $27$& $0.36^{+0.20}_{-0.28}$& 29 \\
 GOODS-S &$-59^{+21}_{-20} $ & $102^{+23}_{-17} $ &$69^{+22}_{-15}$ &$121^{+29}_{-21}$ & $23$ & $28$ & $-0.14^{+0.44}_{-0.72}$& 16\\
 EGS & $-1\pm 10 $ & $72^{+8}_{-7} $ &$59^{+7}_{-6}$ &$139^{+13}_{-11}$ & $20$ & $22$ & $0.77^{+0.05}_{-0.06}$& 66\\

\hline
\end{tabular}
  \caption{Summary of the estimates of the parameters of the halo velocity ellipsoid, for the full sample, the apparent magnitude bins, and the individual HALO7D fields. Posterior medians are quoted, with errorbars giving the 16/84 percentiles.}
\label{tab:results}
\end{center}
\end{table*}

\subsection{Spherically Averaged Estimates}

	We first estimate the parameters of the velocity ellipsoid using the full HALO7D sample of 188 stars. The parameters in this model are the four ellipsoid parameters; the disk contamination in each of the four fields; as well as the absolute magnitude (and therefore distances) to each star. The resulting 1D marginalized distributions for the ellipsoid parameters are shown as the purple histograms in Figure \ref{fig:sphere_sum}. The left-hand panels show histograms of posterior samples for the four halo velocity ellipsoid parameters $\theta_{\rm Halo}=\{\langle v_{\phi} \rangle, \sigma_r, \sigma_{\phi}, \sigma_\theta\}$. Using the full sample of stars, we do not see a strong signature of halo rotation ($\langle v_{\phi} \rangle=-11 \pm 6$ km $s^{-1}$). We use the posterior samples of the ellipsoid parameters to derive a posterior estimate for the velocity anisotropy $\beta$; the resulting posterior distribution is shown as the purple histogram in the lower right-hand panel of Figure \ref{fig:sphere_sum}. We find that $\beta$ is radially biased: at our mean sample distance of $\langle r \rangle=24$ kpc, $\beta=0.63 \pm 0.05$, consistent with estimates of $\beta$ in the solar neighborhood (e.g., \citealt{Bond2010}).
    
    In addition to modeling the full sample of stars, we also split our sample into three apparent magnitude bins. Because our distance estimates to each individual star are uncertain and probabilistic, we cannot divide our sample into different radial ranges; we therefore split the sample in apparent magnitude to study the radial variation of $\beta$. The resulting marginalized posterior distributions for the three apparent magnitude bins are shown as the blue histograms in Figure \ref{fig:sphere_sum}. Estimates using stars with $19.0 < m_{F606W} < 21.0$ are shown in light blue; the estimates from stars with $21.0 < m_{F606W} < 22.5$ are shown as gray blue; and $22.5 < m_{F606W} < 24.5$ are shown in dark blue. The cumulative distribution functions (CDFs) for the distances to each of the three samples are shown in the upper right-hand corner of Figure \ref{fig:sphere_sum}, along with the CDF for distance of the full sample.
    
    Figure \ref{fig:params_v_d} shows the second moments of the velocity distributions as a function of the average Galactocentric distance to the sample. We see a trend of decreasing velocity dispersion with distance, in both tangential and radial motion. However, when we compute the posterior distribution for $\beta$ (blue histograms in lower right-hand panel of Figure \ref{fig:sphere_sum}), we find that all three estimates are consistent with the estimate of $\beta$ from the full sample: $\beta$ is radially biased for all of our spherically-averaged samples. In addition, the posterior medians for $\beta$ increase as a function of mean sample distance, consistent with predictions from simulations (e.g., \citealt{Abadi2006}, \citealt{Loebman2018}).
	
	\subsection{Individual Fields}  
    
    \begin{figure*}
		\centering
		\includegraphics[width=0.8\textwidth]{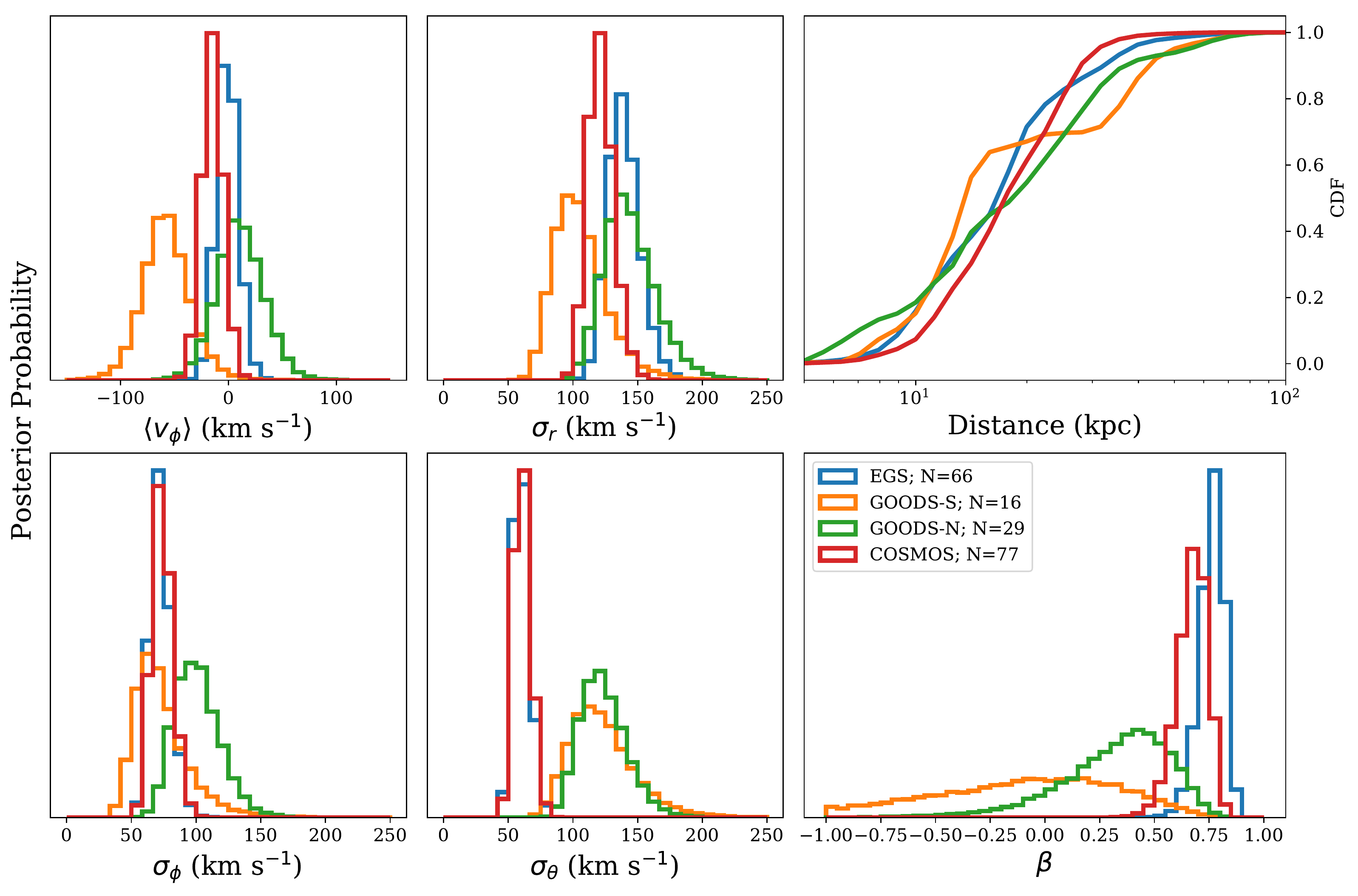}
		\caption{Summary of posterior results for the four fields. Left-hand panels: histograms of marginalized posterior samples for the four parameters of the halo velocity ellipsoid. Each colored histogram represents a different HALO7D field. Upper righthand panel: CDFs for the distances of the stars in the four fields. Lower righthand panel: posterior distributions for $\beta$. }
		\label{fig:four_fields_sum}
	\end{figure*}

	In the case of modeling fields individually, the free parameters in our model include the velocity ellipsoid parameters, the distance to each star in the field, as well as the fraction of disk contamination in the field. 
	
	Posterior samples for the ellipsoid parameters in each of the four fields are shown in Figure \ref{fig:four_fields_sum}. The left-hand panels show histograms of posterior samples for the four halo velocity ellipsoid parameters $\theta_{\rm Halo}={\langle v_{\phi} \rangle, \sigma_r, \sigma_{\phi}, \sigma_\theta}$. The upper right-hand panel shows the cumulative distribution for the distances to the four fields, and the lower right-hand panel shows the resulting posterior distribution for the velocity anisotropy. 
    
    When the four fields are treated separately, we see variation in the estimates of the velocity ellipsoid parameters. While the PDFs for GOODS-S are the broadest, because it has the smallest sample size, the GOODS-S distribution also shows a signature of rotation ($\langle v_{\phi} \rangle =-59^{+21}_{-20}$ km s$^{-1}$). The resulting estimate for $\beta$ is tangentially isotropic, though also very broad, due in part to the small sample size in this field, but also due to the fact that circular orbits correspond to $\beta=-\infty$. In contrast, the estimates in the EGS field show no rotation, and the resulting estimate of $\beta$ is strongly radially biased ($\beta_{\rm EGS} =0.77^{+0.05}_{-0.06}$).
    
    \subsection{Disk Contamination}
    \label{sec:disk_results}
    
    \begin{figure}
    \includegraphics[width=0.5\textwidth]{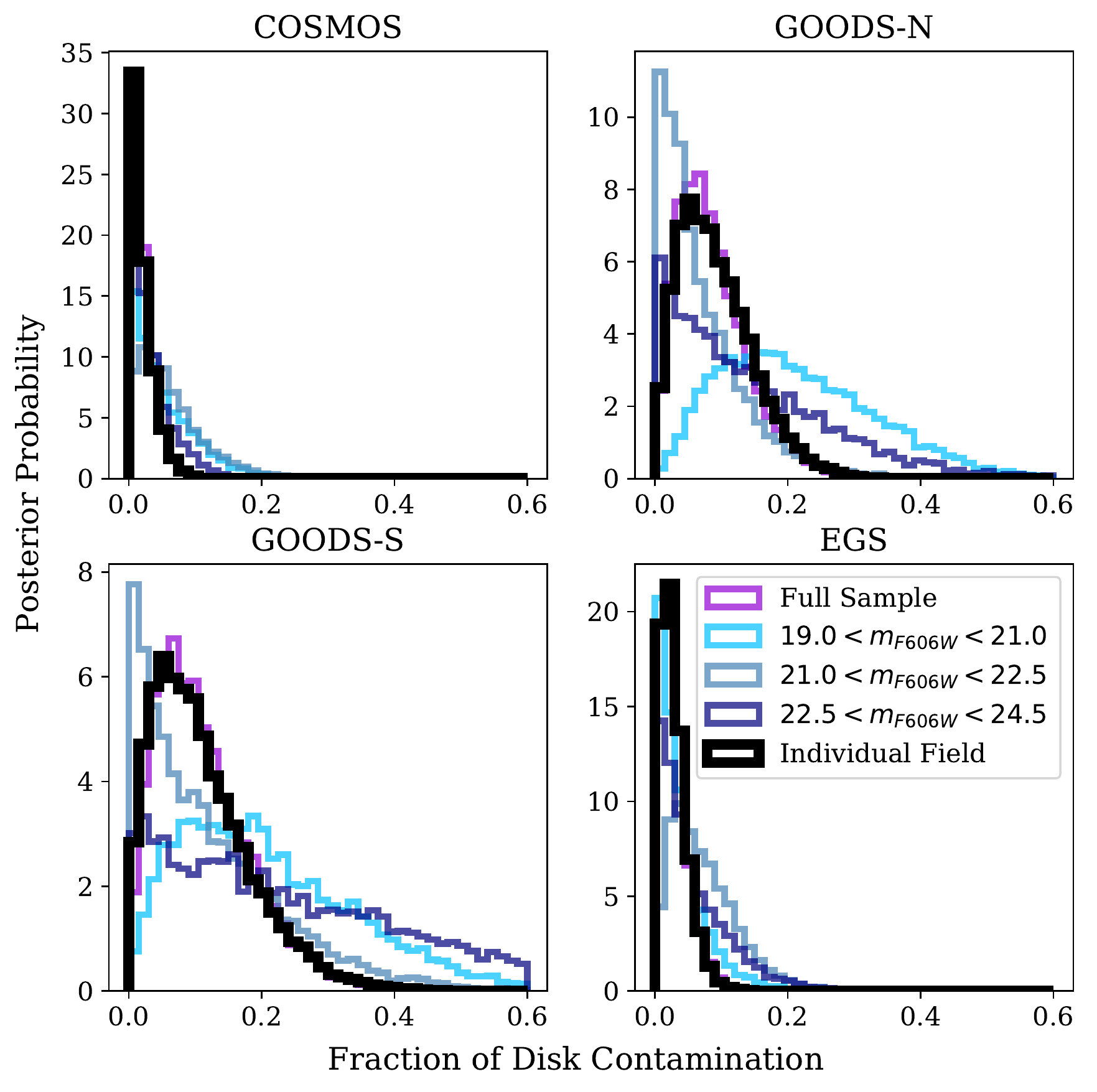}
    \caption{Posterior distributions for the disk contamination in the four HALO7D fields for each of the models. Black histograms indicate the posterior distributions for the fraction of disk stars when the fields are modeled individually. Colored histograms indicate the full sample (purple histograms) and the spherically averaged estimates in different apparent magnitude bins (as in Figures \ref{fig:sphere_sum} and \ref{fig:params_v_d}).}
    \label{fig:disk}
    \end{figure}
    
    The marginalized 1D posterior distributions for the disk contamination in each of the four HALO7D fields are shown in Figure \ref{fig:disk}. The posteriors for $f_{\rm Disk}$ when the fields are treated individually are the thick black histograms; the colored histograms show the estimates in a given field for the spherically averaged estimates. 
    
    Our estimates for disk contamination are low (on the order of or less than $10\%$); this is consistent with the predicted disk contamination levels predicted by the Besan\c{c}on Galaxy Model (\citealt{Robin2003}; see Paper I). Because GOODS-N and GOODS-S have smaller sample sizes than EGS and COSMOS, their posterior distributions for $f_{\rm Disk}$ are broader, but the posterior modes are still around 10\%. As is to be expected, the disk contamination is highest for the brightest apparent magnitude bin (light blue histograms). 
    
    \subsection{The $\beta$ Radial Profile}
    
\begin{figure*}
		\includegraphics[width=\textwidth]{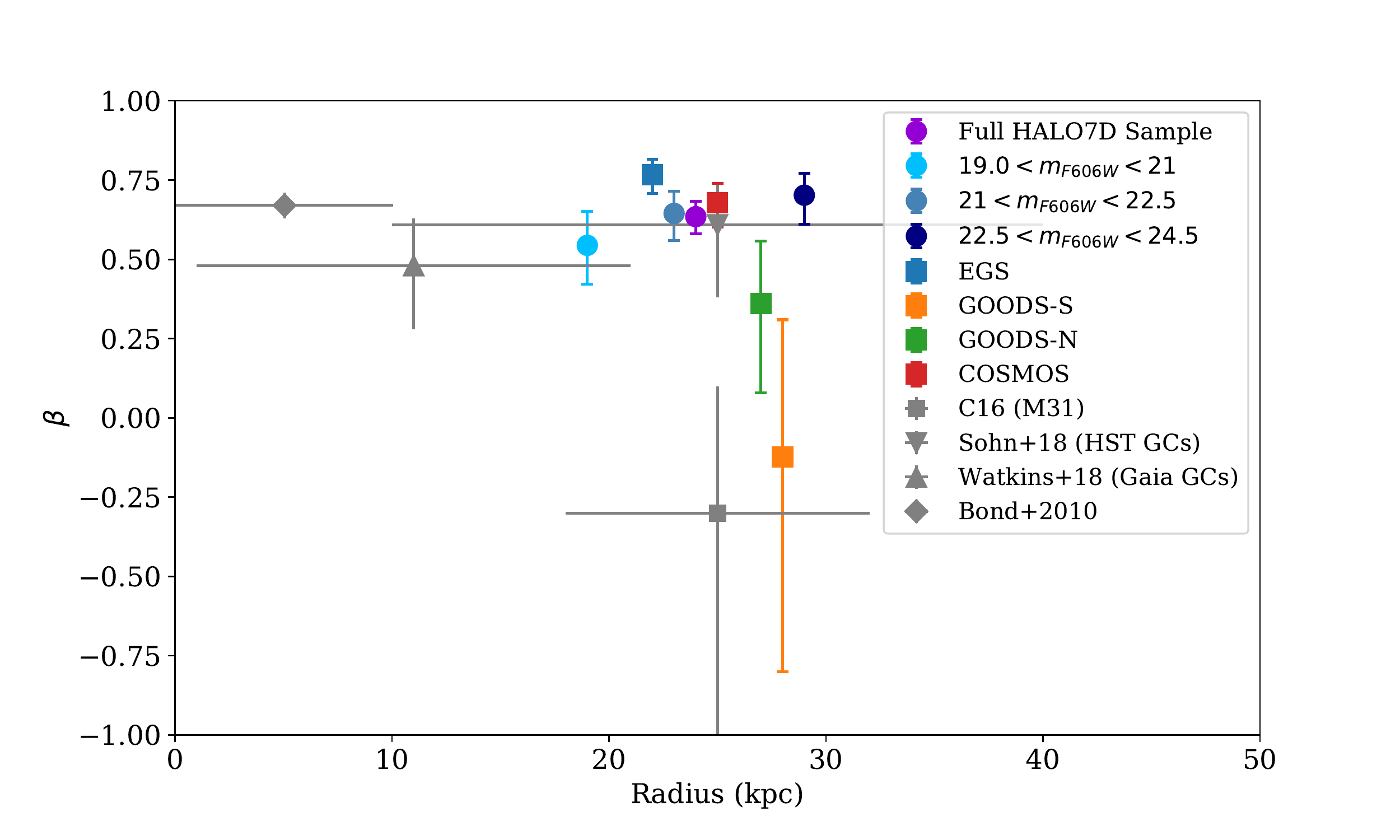}
		\caption{The Milky Way's radial anisotropy profile, $\beta$, as measured from 3D kinematics. Colored points indicate results from this work, while grey points indicate findings from previous work and other studies. The grey diamond shows the anisotropy estimate from \cite{Bond2010}, using main sequence stars from SDSS, and the grey square shows the estimate from \cite{Cunningham2016}, using 13 MW MSTO stars along the line of sight towards M31. The two recent estimates for $\beta$ from MW globular clusters, using HST PMs and \textit{Gaia} PMs, are shown with triangles (\citealt{Sohn2018}, \citealt{Watkins2018}). Square shaped points are results from individual fields, while our spherically averaged results are shown as circles. When using small fields to estimate anisotropy, $\beta$ varies from mildly tangential (e.g. GOODS-S, M31) to strongly radial (e.g., EGS). However, the spherically averaged estimates are all consistently $\beta \sim 0.6$ (and consistent with solar neighborhood and GC estimates), and the posterior means increase as a function of mean sample distance.}
		\label{fig:beta_v_d_full}
	\end{figure*}

    Figure \ref{fig:beta_v_d_full} summarizes all of our estimates of the velocity anisotropy, as a function of mean distance. Our spherically averaged estimates are plotted as circles, while the estimates of our individual fields are shown as squares. Gray points are results from other studies that used 3D kinematics to estimate $\beta$: gray triangles show the estimates of $\beta$ from MW GCs, using PMs from both HST (\citealt{Sohn2018}) and \textit{Gaia} (\citealt{Watkins2018}). The gray diamond shows the 3D estimate of $\beta$ in the solar neighborhood from SDSS (\citealt{Bond2010}), and the gray square shows the C16 estimate of $\beta$ along the line of sight towards M31. 
    
    Our spherically averaged estimates of $\beta$, which find radially biased $\beta \sim 0.6$, are consistent with one another and with other studies that have estimated $\beta$ averaging over different parts of the sky. However, our field-to-field estimates (including the estimate from C16) show substantial variation, from strongly radially biased (EGS) to mildly tangentially biased (GOODS-S; M31). While the GOODS-S and M31 fields each have lower posterior estimates for $\beta$, these two fields also have the smallest sample size. Because of the way $\beta$ is defined, estimates of $\beta$ are sensitive to sample size and measurement uncertainties. We therefore assess how much our sample size should concern us by testing fake data. We generate 100 fake datasets (in the method described in Appendix \ref{sec:fake_data_ellipsoid}), from velocity distributions that have $\beta_{\rm True}=0.75$, each containing 16 stars. Out of the 100 tests, only one fake dataset had a posterior distribution for $\beta$ with median $\beta_{\rm Med}\leq 0$ (for the full distribution of $\beta$ posterior medians, see Figure \ref{fig:beta_fake_goodss} in Appendix \ref{sec:fake_data_ellipsoid}). We do not see small sample size resulting in a systematic underestimate of $\beta$. Therefore, while the small sample size does contribute to the large uncertainty on $\beta$ in this field, based on our fake data testing, we do not expect that the observed mildly tangential $\beta$ is purely due to sample size.

	\section{Comparison with Other Studies}
	\label{sec:disc}
    
    In this paper, we use the HALO7D dataset to estimate the parameters of the MW stellar halo velocity ellipsoid. We study the full HALO7D sample, the sample divided into three apparent magnitude bins, and the individual HALO7D fields. When averaging over the four HALO7D fields, we find consistent estimates for $\beta \sim 0.6$, with posterior medians increasing as a function of mean sample distance. Our spherically averaged results for $\beta$ are consistent with results from other recent estimates of $\beta$ using GCs as tracers (\citealt{Sohn2018}, \citealt{Watkins2018}). However, when we treat the four HALO7D fields separately, our estimates for the ellipsoid parameters, and thus $\beta$, show significant variation.
   
    In their study of $\beta$ profiles of simulated galaxies, \cite{Loebman2018} found that $\beta$ profiles are generally increasingly radially biased as a function of radius. However, recently accreted material can cause short-lived ($<0.2$ Gyr) dips in the $\beta$ profiles, and longer-lived  ($>0.2$ Gyr) dips arise due to the disruption of the in-situ stellar halo by the close passage of a massive satellite. These ``dips'' in the in-situ stellar halo are more metal-rich than dips caused by the accreted stellar halo. 
    
    Several studies using LOS velocities alone have observed tangentially biased ``dips" in the $\beta$ profile (\citealt{Sirko2004}; \citealt{Kafle2012}, \citealt{King2015}); these dips occur approximately at the observed break in the MW density profile (\citealt{Deason2011b}, \citealt{Sesar2011}, \citealt{Watkins2009}). The kinematic structure around the break radius $r_b \sim 27$ kpc is of interest in order to understand its origin. In their study of the \cite{Bullock2005} purely accreted stellar halos, \cite{Deason2013a} found that the buildup of stars at apocenter from a relatively early, massive accretion event, or a few synchronous events, can cause broken density profiles. 
    
    As pointed out by \cite{Hattori2017}, studies of stars beyond $r\sim 15$ kpc using only LOS velocities are subject to underestimates of $\beta$. However, recent studies using \textit{Gaia} PMs have found decreases in $\beta$ around the break radius as well. Using blue horizontal branch stars in \textit{Gaia} DR2, \cite{Lancaster2018} found that $\beta$ decreases just beyond the break radius, from $\beta\sim0.6$ at 20 kpc to $\beta\sim0.4$ at 40 kpc. They argue that this is due to sharp decline in the fraction of stars belonging to a radially biased population that dominates the inner halo (i.e., the \textit{Gaia}-Sausage) beyond its apocenter radius (which \citealt{Deason2018b} showed coincides with the MW break radius). Using LAMOST K-Giants with \textit{Gaia} DR2 PMs, \cite{Bird2018} found strongly radially biased ($\beta\sim 0.8$) inside of $r\sim25$ kpc, with $\beta$ gradually decreasing beyond this radius, down to $\beta=0.3$ at 100 kpc; however, \cite{Lancaster2018} also showed that the magnitude of the decrease observed in the \cite{Bird2018} study could be due to their treatment of measurement uncertainties. 
    
    When averaging over multiple fields, we do not see a dip in the $\beta$ profile, nor a global decrease in $\beta$ beyond the break radius. While our estimates are around the MW break radius, the posterior medians of our spherically averaged estimates increase as a function of Galactocentric distance. 
Increasing $\beta$ as a function of radius is consistent with predictions from simulations (e.g., \citealt{Abadi2006}, \citealt{Sales2007}, \citealt{Rashkov2013}, \citealt{Loebman2018}). However, we need to probe to larger distances beyond the break radius to see if this trend continues to larger radii, or if $\beta$ starts to decrease (as seen by \citealt{Bird2018} and \citealt{Lancaster2018}).
    
When we treat our different lines-of-sight separately, we see potential evidence for a dip in $\beta$ towards GOODS-S and M31. Based on the \cite{Loebman2018} findings, these sightlines could be dominated by material that has been recently accreted or kicked up by the passage of Sagittarius. As discussed in the Introduction, several overdensities previously believed to be accreted structures now show evidence of a potential disk origin, having been kicked out of the disk due to the passage of Sagittarius (e.g.,\citealt{Price-Whelan2015}; \citealt{Laporte2018}; \citealt{Bergemann2018}). One such overdensity discussed in those works is TriAnd, located along the line of sight towards M31, which is also the lowest latitude of the HALO7D fields.
   
    Measuring abundances for stars in the HALO7D fields from their Keck spectra (McKinnon et al., in prep) will help to distinguish between the kicked-up disk scenario and the recent accretion scenario as the origin for the observed ``dips" in $\beta$ in GOODS-S and M31. Chemical abundances will also help to assess the origin of the strongly radially biased $\beta$ estimate in EGS ($\beta_{\rm EGS} \sim 0.8$).  \cite{Belokurov2018} discovered the ``\textit{Gaia}-Sausage" as a metal-rich ($[\mathrm{Fe/H}]>-1.7$), radially biased ($\beta \sim 0.9$) population in \textit{Gaia} DR1. Given that the estimate of $\beta$ in EGS is more radially biased than the estimates of $\beta$ in the other fields, it is possible that the sample of stars in EGS is dominated by Sausage stars. Chemical abundances will be essential in assessing to what extent the Sausage is contributing to the HALO7D sample. 
    
        \begin{figure*}
		\includegraphics[width=\textwidth]{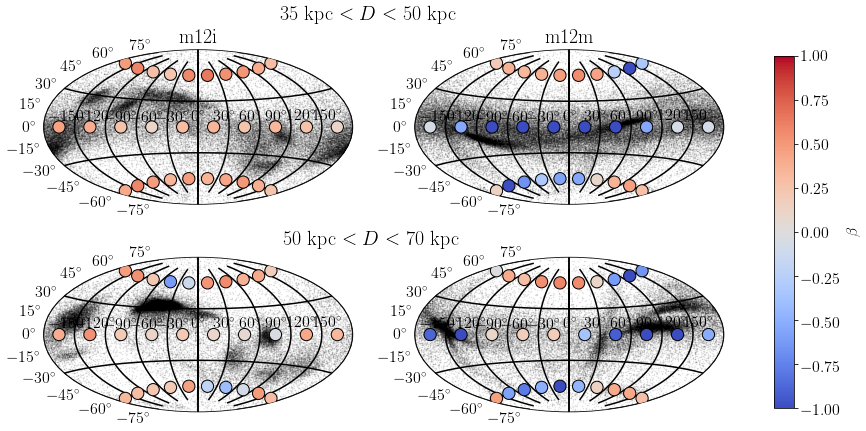}
		\caption{Maps of velocity anisotropy of the stellar halo in two \textit{Latte} FIRE-2 simulations of Milky Way-mass galaxies: m12i (left) and m12m (right). The top panels show stars in the distance range $35~\mathrm{kpc} < D < 50~\mathrm{kpc}$, while the lower panels show $50~\mathrm{kpc} < D < 70~\mathrm{kpc}$. Black points show the positions of star particles in Galactic coordinates. In each halo, the sky has been subdivided into patches, and the color of the large circle in each patch indicates the velocity anisotropy for that subset of stars. Within a given distance range, each halo shows variation in its velocity anisotropy across the sky. Variation as a function of distance is also evident. In addition, the median and spread in $\beta$ both vary from halo to halo: many more of the star particles in m12m are on tangentially biased orbits than in m12i.}
		\label{fig:beta_maps}
\end{figure*}
    
    \section{Comparison with Simulations}
    \label{sec:latte}
    
    When we treat the four HALO7D fields separately, we see variation in the estimates of the velocity ellipsoid parameters (and the resulting velocity anisotropy). In this section, we explore the spatial variation of velocity anisotropy in two halos from the \textit{Latte} suite of FIRE-2 cosmological zoom-in baryonic simulations of Milky Way-mass galaxies (introduced in \citealt{Wetzel2016}), part of the Feedback In Realistic Environments (FIRE) simulation project.\footnote{FIRE project website: \url{http://fire.northwestern.edu}} These simulations were run using the Gizmo gravity plus hydrodynamics code in meshless finite-mass (MFM) mode (\citealt{Hopkins2015}) and the FIRE-2 physics model (\citealt{Hopkins2018}). In this work, we discuss halos m12i (initially presented in \citealt{Wetzel2016}) and m12m (initially presented in \citealt{Hopkins2018}), making use of the publicly available $z=0$ snapshots (\citealt{Sanderson2018}).\footnote{$z=0$ snapshots available at \url{http://ananke.hub.yt}} The properties of the halos of these galaxies have been shown to agree reasonably well with the properties of the MW and M31, including the stellar-to-halo mass relation (\citealt{Hopkins2018}); satellite dwarf galaxy stellar masses, stellar velocity dispersion, metallicities, and star-formation histories (\citealt{Wetzel2016}, \citealt{Garrison-Kimmel2018}, \citealt{Escala2018}, Garrison-Kimmel et al., in prep); and stellar halos (\citealt{Sanderson2017}, \citealt{Bonaca2017}).
In particular, the high resolution of the \textit{Latte} simulations (star particles have initial masses $m \approx 7000 M_{\odot}$ and gravitational force softening of 4 pc) means that they resolve satellite dwarf galaxies down to $M_{\rm star} \gtrsim 10^5 M_\odot$, thus resolving the galaxies that are expected to contribute the majority of mass to the formation of the stellar halos (e.g., \citealt{Deason2015}).

Despite the high resolution of the \textit{Latte} simulations, at large distances in the halos the typical spacing between star particles can be large compared to the sizes of the HALO7D fields (less than a square degree). Therefore, a detailed comparison of exactly how our selection effects, observational errors and field sizes are affecting our results is beyond the scope of this work. In addition, such a comparison may be of limited usefulness, given that variation observed across areas as small as the HALO7D fields could be due to structures below the simulation resolution limit (i.e., debris from accreted satellites with $M_{\rm star} < 10^5 M_\odot$). As a first step, we explore the spatial variation in the velocity anisotropy computed directly from the star particles in the simulation, using 30 larger fields, each spanning 36 degrees in longitude and 60 degrees in latitude.

Figure \ref{fig:beta_maps} shows the m12i simulation (left panels) and m12m (right panels). The top panels show the positions of star particles (black points), in Aitoff projection, within the distance range $35<D<50$ kpc; the lower panels show star particles in the distance range $50<D<70$ kpc. While these radial ranges are farther out than the HALO7D data, we choose these ranges to avoid the thick disks in these simulations, which are extended and kinematically hotter than the MW (\citealt{Sanderson2018}, Loebman et al, in prep). Star particle positions are plotted in Galactic coordinates. Galactocentric frames are defined in the method described in Section 3 of \cite{Sanderson2018}, and positions are converted to Galactic coordinates using the \verb+astropy.coordinates+ package (\citealt{Astropy2013}; \citealt{Astropy2018}). We use the default options in \verb+astropy.coordinates+ for the position of the Sun ($R_{\odot}=8.3$ kpc; \citealt{Gillessen2009}). The ``sky'' in each halo has been divided into patches, and the resulting velocity anisotropy computed from the star particles in each patch is shown by the colored points. We note that we compute the velocity anisotropy using all the star particles within a given area on the sky and radial range; we do not exclude particles in bound satellite galaxies. This choice likely affects the resulting $\beta$ maps, and we plan to explore the affects of excluding and including bound satellites on $\beta$ estimates in future work.

The velocity anisotropy $\beta$ shows variation across the sky of a given halo, as well as with radius. In addition, these maps of velocity anisotropy are very different across the two halos: within $35<D<50$ kpc all patches in m12i are radially biased, but many of the patches of m12m are tangentially biased $\beta$. At $50<D<70$ kpc, both m12i and m12m show some tangentially biased patches and radially biased patches. The overall means and spreads of $\beta$ values measured across the two halos are quite different: m12i has a mildly radially biased $\langle \beta \rangle \sim 0.3$ with a standard deviation of 0.2, whereas m12m has tangentially biased $\langle \beta \rangle \sim -0.3$ with a standard deviation of 0.7. The magnitude of the $\beta$ variation observed in these two simulated galaxies is very similar to the range of $\beta$ values measured across the HALO7D fields; in both Figures \ref{fig:beta_v_d_full} and \ref{fig:beta_maps}, we see variation in $\beta$ over the range of $[-1,1]$. 

The differences in the $\beta$ maps across these two halos are likely linked to their different accretion histories. Over these radial ranges in the simulated halos, the majority of the material mapped in Figure \ref{fig:beta_maps} is accreted, and the accreted debris in the two halos have visibly different spatial and kinematic properties. We intend to explore in future work what characteristics of a galaxy's accretion history, such as accretion times, initial orbital conditions, and masses of accreted satellites, are primarily responsible for the observed $\beta$ variation patterns. Based on the \cite{Loebman2018} findings, patches with tangentially biased $\beta$ could be indicating recently accreted material. Further study of the accretion histories of these simulated galaxies will help us to understand what accretion events and accretion histories give rise to different $\beta$ variation patterns in galaxy halos, and what characteristics of the MW's assembly history we might be able to constrain through mapping its spatial $\beta$ variation. 

The $\beta$ variation we observe in HALO7D and the \textit{Latte} simulations also could have implications for the validity of MW mass estimates derived from Jeans modeling. The fundamental assumption underlying Jeans modeling is that the tracers are virialized and in dynamical equilibrium. The spatial maps and $\beta$ variation observed in the Latte halos reveal that this assumption is clearly violated in the simulations. The variation in $\beta$ observed with HALO7D indicates that this assumption is invalid in the MW halo as well; our results are evidence that the halo is not phase-mixed at $\langle r \rangle\sim 24$ kpc. Just how significantly the violation of the assumption of dynamical equilibrium will affect estimates of the MW mass remains to be determined. The systematic uncertainty of traditional spherical Jeans mass modeling in recovering halo masses has been observed in a number of simulations (e.g., \citealt{Wang2018}, \citealt{Kafle2018}, \citealt{Eadie2018}); we leave the full characterization of the effects of $\beta$ variation on different approaches of MW mass estimates on the \textit{Latte} halos to future work. 
	
	\section{Conclusions}
	\label{sec:concl}
	
	In this paper, we present the proper motions of distant, main sequence turnoff MW halo star candidates as measured with \textit{HST}. These PMs are measured as a part of the HALO7D project, and have LOS velocities measured from Keck spectroscopy (Paper I). Using the 3D kinematic sample from HALO7D, we estimate the parameters of the halo velocity ellipsoid and velocity anisotropy. We estimate these parameters treating the individual survey fields separately as well as spherically averaging over all fields.
    
    We summarize our main results as follows:
    \begin{enumerate}
    \item Using the full HALO7D sample of 188 stars, we estimate the velocity anisotropy $\beta=0.63 \pm 0.05$ at $\langle r \rangle = 24$ kpc. This estimate is consistent with other recent estimates of $\beta$.
    \item We estimate $\beta$ from the HALO7D sample split into three apparent magnitude bins to explore the radial dependence. While estimates of velocity dispersions decrease as a function of mean sample distance, the overall estimates of $\beta$ are consistent across apparent magnitude bins. Posterior medians increase as a function of mean sample distance, consistent with predictions from simulations.
    \item When we treat our stars from the four HALO7D fields separately, estimates of the halo velocity ellipsoid parameters show variation from field to field. This variation could be evidence for recent accretion; it is also possible that the tangentially biased $\beta$ values from GOODS-S and M31 are due to the presence of kicked-up disk stars. The observed variation in $\beta$ is evidence that the halo is not phase-mixed at $r \sim 24$ kpc.
    \item We map the velocity anisotropy in two stellar halos from the \textit{Latte} suite of FIRE-2 simulations and see variation in $\beta$ across the skies of these two halos over a similar range to the variations observed in the HALO7D fields. In the simulated galaxies, the degree of, and patterns in, these variations are clearly tied to their different accretion histories. A more detailed study of the full accretion histories of these galaxies will shed light on the types of signatures that different accretion events can leave in $\beta$ maps.

	\end{enumerate}
    
    Fortunately, many of the questions raised in this work are answerable in the near future. Abundances from HALO7D spectra will provide key insights as to the origin of the observed $\beta$ variation. In addition, $\beta$ variation in the MW can be mapped with the \textit{Gaia} dataset and, ultimately, LSST. Thanks to the quality of current and upcoming data, coupled with high resolution cosmological hydrodynamical simulations, we are rapidly progressing in our knowledge of our Galaxy's structure and formation.

\section*{Acknowledgments}
	Over the course of this work, ECC was supported by a NSF Graduate Research Fellowship as well as NSF Grant AST-1616540. Partial support for this work was provided
by NASA through grants for program AR-13272 from the
Space Telescope Science Institute (STScI), which is operated
by the Association of Universities for Research in Astronomy
(AURA), Inc., under NASA contract NAS5-26555. A.D. is supported by a Royal Society University Research Fellowship.
A.D. also acknowledges support from the STFC grant ST/P000451/1.
AW was supported by NASA through ATP grant 80NSSC18K1097 and grants HST-GO-14734 and HST-AR-15057 from STScI.
SL was supported by NASA through Hubble Fellowship grant \#HST-JF2-51395.001-A awarded by the Space Telescope Science Institute, which is operated by the Association of Universities for Research in Astronomy, Inc., for NASA, under contract NAS5-26555.
This project was carried out in
the context of, and used software created by, the HSTPROMO
(High-resolution Space Telescope PROper MOtion)
Collaboration.\footnote{http://www.stsci.edu/~marel/hstpromo.html} This research made use of Astropy, a community-developed core Python package for Astronomy (\citealt{Astropy2013}; \citealt{Astropy2018}). This work has made use of the Rainbow Cosmological Surveys Database, which is operated by the Universidad Complutense de Madrid (UCM), partnered with the University of California Observatories at Santa Cruz (UCO/Lick,UCSC). We recognize and acknowledge the significant cultural role and reverence that the summit of Mauna Kea has always had within the indigenous Hawaiian community. We are most fortunate to have the opportunity to conduct observations from this mountain.

	\appendix
	
	In this appendix, we provide the details of the Bayesian mixture model used to derive estimates of PMs. We first describe our model in Section \ref{sec:pm_model}, and then the Gibbs sampling algorithm used to sample from the posterior distribution for model parameters in Section \ref{sec:gibbs}.
	
		\section{Proper Motion Model}
		\label{sec:pm_model}
		
		To map one image onto another, we perform a six-parameter linear transformation:
	
		\begin{equation}
			\begin{pmatrix}
				A && B && x_t-\delta u \\
				C && D && y_t -\delta v\\
				0 && 0 && 1   \\
			\end{pmatrix}
			\quad
			\begin{pmatrix}
				&u& \\
				&v& \\
				&1& \\
			\end{pmatrix}
			=
			\begin{pmatrix}
				u_{ref} \\
				v_{ref} \\
				1 \\
			\end{pmatrix},	
		\end{equation}
where $\delta u, \delta v$ represent the change (in pixels) of a star from one image to another due to proper motion (so, for galaxies, $\delta u = \delta v = 0$). 
	
		In our model, we treat stars and galaxies separately. For stars:
		\begin{equation}
		\begin{split}
			Au+Bv+x_t-\delta u \sim \mathrm{N}(u_{ref}, \sigma_{*}^2) \\
			Cu+Dv+y_t-\delta v \sim \mathrm{N}(v_{ref}, \sigma_{*}^2) \\
			u_{im ref}, v_{im ref} \sim \mathrm{N}(u_{ref}, v_{ref}, \sigma_{*}^2)
		\end{split},
		\end{equation}
where $u_{im ref}, v_{im ref}$ are the measured positions in the defined reference image, whereas $u_{ref}, v_{ref}$ are the positions of the object in the reference epoch, which are free parameters. Because stars' central positions are well measured, we define $\sigma_{*}=0.02$ pixels. $\delta u, \delta v$ represent the shift in pixels from image 1 to image 2, which can be converted to proper motions North and West, respectively:
	
		\begin{equation}
		PMN = \frac{\delta u \times 50~\mathrm{mas}/\mathrm{pix}}{\Delta t};	
		\end{equation}
	
		\begin{equation}
		PMW = \frac{\delta v \times 50~\mathrm{mas}/\mathrm{pix}}{\Delta t},	
		\end{equation}
where $\Delta t$ is measured in years. We describe the galaxy positions as a two-component Gaussian mixture model. Defining a fixed location in an image as the galaxy's precise position is not trivial and sometimes fails, given that galaxies are resolved sources with complex morphologies. We therefore consider ``good" galaxies (i.e., galaxies with well measured positions) and ``bad" galaxies (galaxies with poorly measured positions).  
	
		\begin{equation}
		\begin{split}
			Au+Bv+x_t \sim \mathrm{N}(u_{ref}, \sigma^2) \\
			Cu+Dv+y_t \sim \mathrm{N}(v_{ref}, \sigma^2) \\
			u_{imref}, v_{imref} \sim \mathrm{N}(u_{ref}, v_{ref}, \sigma^2) \\
		\end{split}.
		\end{equation}
	
		For ``good" galaxies, $\sigma=0.1$ pixels, whereas for ``bad" galaxies, $\sigma=3$ pixels. 
        
\begin{figure}
        \centering
        \includegraphics[width=0.5\textwidth]{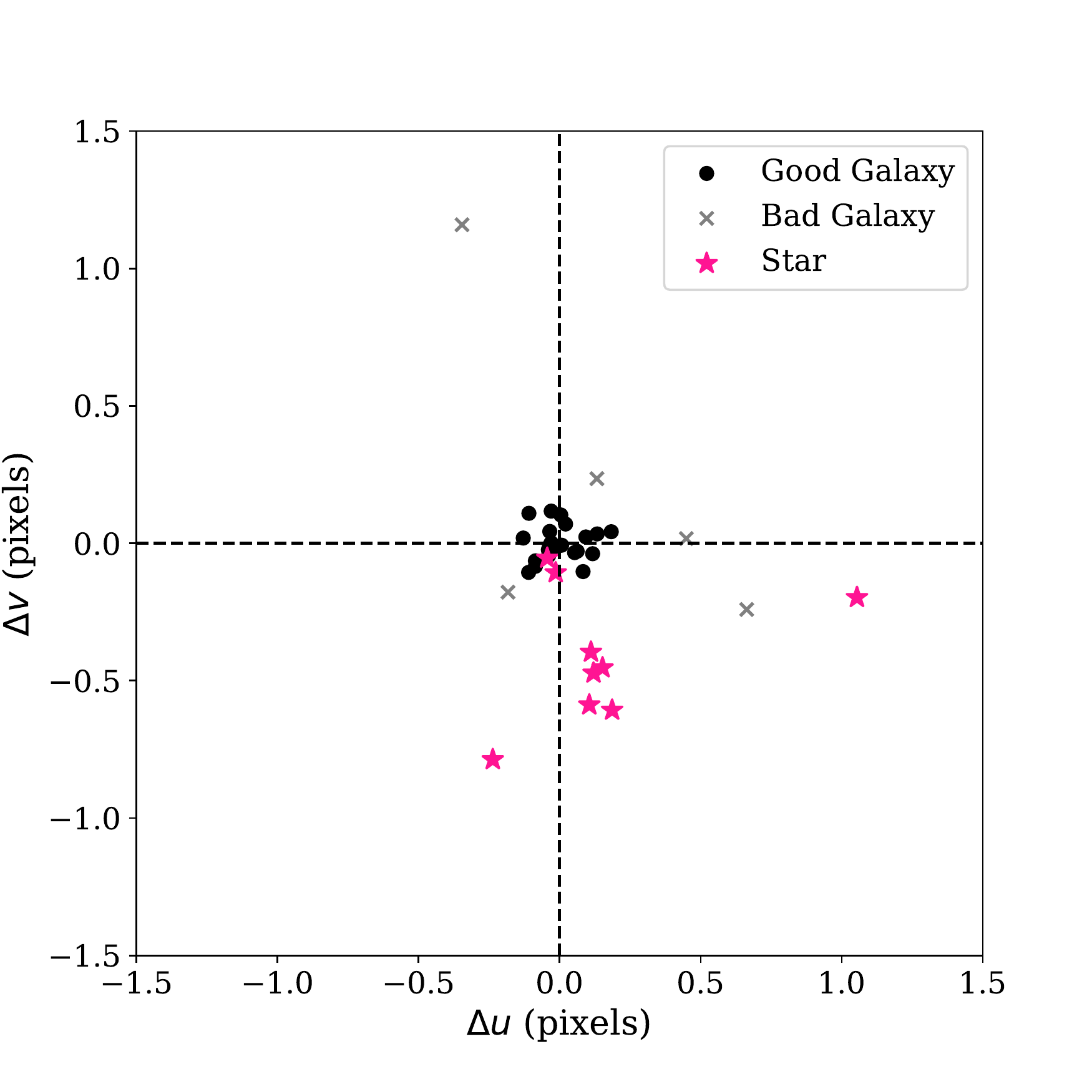}
        \caption{An example of the linear transformation method on two images, j8pu44cvq (taken in 2004) and jboa38c2q (from 2011). Axes represent the change in pixels, in the distortion-corrected frame $(u,v)$, for objects in the two images, after applying the six parameter linear transformation. Grey crosses indicate the change in positions for the galaxies initially classified as ``bad''; black points are the positions of ``good'' galaxies used in the reference frame. The change in positions of the ``good'' galaxies are clustered at $(0,0)$. Pink stars show change in the positions of the stars in these two images.}
        \label{fig:pm_trans}
        \end{figure}

    An example of the initial classification of ``good" and ``bad" galaxies is shown in Figure \ref{fig:pm_trans}. Figure \ref{fig:pm_trans} shows the change in positions in pixels, in the distortion-corrected frame, for objects in two \textit{HST} images, taken seven years apart. Black points show the positions of galaxies initially classified as ``good"; these are clustered at $(0,0)$, because they were used in the reference frame for the linear transformation. Positions of galaxies initially classified as ``bad" are shown as grey crosses. In our Bayesian mixture model, we allow galaxies to move in and out of the reference frame probabilistically. Pink points show the change in the positions of the stars in the images. These stars have a mean motion and scatter relative to the stationary reference frame of distant galaxies; these relate to the dynamical quantities of interest estimated in this study.    
        
		\subsection{Gaussian Mixture Models written with Indicator Variables}
	
		Mixture models can be expressed in different ways. For a two component mixture model, the likelihood of a given data point can be written as
	
		\begin{equation}
			p(x|\theta) = \lambda N(x|\theta_1, \sigma_1^2) + (1-\lambda) N(x|\theta_2, \sigma_2^2)
		\end{equation}
	
		Where $\lambda$ is the fraction of objects in the underlying population that belong to distribution 1. However, sums in probability calculations make posterior sampling more difficult. To improve our sampling efficiency, we can re-write the above equation using indicators $z_j$:
	
		\begin{equation}
			p(x, z|\theta) = (\lambda N(x|\theta_1, \sigma_1^2))^{z_{1}}((1-\lambda) N(x|\theta_2, \sigma_2^2))^{z_2}.
		\end{equation}
In this construction, for a given step in the MCMC chain, the indicator $z_{ih}=1$ if datapoint $x_i$ is associated with component $h$  and $z_{ih}=0$ otherwise.
		
		Our full posterior thus takes the form:
	
		\begin{equation}
			\begin{split}
		p(\theta|u_1...u_k,v_1...v_k,u_{imref},v_{imref}) \propto  \\
		 \prod_{k=1}^{N_{im}} \prod_{j=1}^{N_{stars}} \exp \left \{\frac{1}{2\sigma_*^2}(Au_{jk}+Bv_{jk}+x_{t,k}-\delta u_{jk}-u_{ref})^2 + \frac{1}{2\sigma_*^2}(Cu_{jk}+Dv_{jk}+y_{t,k}-\delta v_{jk}-v_{ref})^2 \right \} \\
		\times \prod_{j=1}^{N_{gals}} \left( f_{g,k} \exp \left \{\frac{1}{2\sigma_g^2}(Au_{jk}+Bv_{jk}+x_{t,k}-u_{ref})^2 + \frac{1}{2\sigma_*^2}(Cu_{jk}+Dv_{jk}+y_{t,k}-v_{ref})^2 \right \}\right) ^{z_{g, jk}} \\
		 \times \left( (1-f_{g,k}) \exp \left \{ \frac{1}{2\sigma_b^2}(Au_{jk}+Bv_{jk}+x_{t,k}-u_{ref})^2 +\frac{1}{2\sigma_b^2}(Cu_{jk}+Dv_{jk}+y_{t,k}-v_{ref})^2 \right \} \right) ^{z_{b, jk}}\\
			\end{split}
		\end{equation}
	
		where $f_{g,k}$ is the fraction of good galaxies in image $k$, and $z_{jg}$ is the indicator for galaxy $j$ in image $k$. By construction, if a galaxy has a ``good'' position in image $k$, $z_{g, jk}=1$ and $z_{b, jk}=0$ (i.e. a galaxy can only belong to one mixture component at a time). 
	
		\subsection{Gibbs Sampling Algorithm}
		\label{sec:gibbs}
	
		To sample from the posterior distribution for our parameters, we use Gibbs sampling. In a Gibbs sampler, we sample directly from the conditional posterior distributions for each parameter. Gibbs samplers can only be used if the full conditional distributions of the parameters can be written in closed form, which is usually only the case when conjugate priors (or, in special cases, reference priors) have been used. 
	
		Our Gibbs sampling algorithm consists of the following steps:
		\begin{enumerate}
		\item Initialize the transformation parameters for each image using standard linear-least squares. If the image is in the same epoch as the reference image, use the star positions to match frames. Otherwise, use the positions of the ``good" galaxies. Initial values for PMs are averaged over the images, and initial values for the reference positions are those in the reference image. 
		\item For each star, we draw from the conditional posterior distributions for PMN and PMW, as well as the conditional posterior distributions for the reference positions. The conditional distributions for proper motions are:
	
		\begin{equation}
			PMW \sim N(\frac{50~\mathrm{mas}/\mathrm{pix}}{k \sum \Delta t_k^2} \sum_{k=1}^{N_{im,k}} \delta u_k \times \Delta t_k, (50 ~\mathrm{mas}/\mathrm{pix})^2 \times \frac{\sigma_{*}^2}{\sum \Delta t^2})
		\end{equation}
	
		\begin{equation}
			PMN \sim N(\frac{50~\mathrm{mas}/\mathrm{pix}}{k \sum \Delta t_k^2} \sum_{k=1}^{N_{im,k}} \delta v_k \times \Delta t_k, (50 ~\mathrm{mas}/\mathrm{pix})^2 \times \frac{\sigma_{*}^2}{\sum \Delta t^2})
		\end{equation}
        
		\item For each galaxy, we first loop over each image, including the reference image, and draw an indicator. We allow a galaxy to be ``good'' in some subset of images and ``bad'' in another. We draw the indicator for a given galaxy as a Bernoulli variable with probability:

	\begin{equation}
		p = \frac{f_{gal}\times N(u|u_{ref}, \sigma_g^2)\times N(v|v_{ref}, \sigma_g^2)}{f_{gal}\times N(u|u_{ref}, \sigma_g^2)\times N(v|v_{ref}, \sigma_g^2)+(1-f_{gal})\times N(u|u_{ref}, \sigma_b^2)\times N(v|v_{ref}, \sigma_b^2)}
	\end{equation}
	
		We then draw from the conditional posterior distributions for the reference frame positions of each object.
	
		\item Finally, we loop over each image, drawing from the conditional posterior distributions for the image parameters $(A,B,C,D, u_t, y_t, f_{gal})$ (see Table \ref{tab:cond_params}). The $f_{gal}$ parameters are drawn from beta distributions:
	
		\begin{equation}
			p(f_{gal,k}|...) \sim \mathrm{Beta}(n_{good gals,k}+1, n_{bad gals,k}+1)
		\end{equation}
	
		where $n_{good gals,k},n_{bad gals,k}$ are computed by summing the indicators for each population at that step in the chain. 
		\end{enumerate}

		\begin{table}
			\begin{centering}
			\begin{tabular}{c c  c }
			Parameter & $\mu/V$ & $V^{-1}$ \\
			\hline
			$A$ &  $\sum \frac{u_j \times u_{ref, j}}{\sigma_j^2} - B \sum \frac{u_j \times v_{j}}{\sigma_j^2} - x_t \sum \frac{u_j}{\sigma_j^2} + \sum \frac{\delta u_{j}u_j}{\sigma_j^2} $ & $\sum \frac{u_j^2}{\sigma_j^2} $\\
			$B$ & $ \sum \frac{v_j \times u_{ref, j}}{\sigma_j^2} - A \sum \frac{u_j \times v_{j}}{\sigma_j^2} - x_t \sum \frac{v_j}{\sigma_j^2} + \sum \frac{\delta u_{j}v_j}{\sigma_j^2} $ & $ \sum \frac{v_j^2}{\sigma_j^2} $ \\
			$C$ & $ \sum \frac{u_j \times v_{ref, j}}{\sigma_j^2} - D \sum \frac{u_j \times v_{j}}{\sigma_j^2} - y_t \sum \frac{u_j}{\sigma_j^2} + \sum \frac{\delta v_{j}u_j}{\sigma_j^2} $ &$  \sum \frac{u_j^2}{\sigma_j^2} $ \\
			$D$ & $ \sum \frac{v_j \times v_{ref, j}}{\sigma_j^2} - C \sum \frac{u_j \times v_{j}}{\sigma_j^2} - y_t \sum \frac{u_j}{\sigma_j^2} + \sum \frac{\delta v_{j}v_j}{\sigma_j^2} $&$ \sum \frac{v_j^2}{\sigma_j^2}$ \\
			$x_{t}$ &$ A \sum \frac{u_j}{\sigma_j^2} + B \sum \frac{v_j}{\sigma_j^2} $ & $ n_{obj}/ \sigma_j^2 $\\
			$y_t$ &$ C \sum \frac{u_j}{\sigma_j^2} + D \sum \frac{v_j}{\sigma_j^2} $ & $ n_{obj}/ \sigma_j^2 $\\

			\end{tabular}
			\caption{Parameters of the conditional posterior distributions for the image transformation parameters. Conditional posteriors for all 6 image transformation parameters are normal distributions with mean $\mu$ and variance $V$. Sums are over all objects in an image. $\sigma_j$=0.02 for stars,$\sigma_j$=0.1 for ``good'' galaxies, and $\sigma_j$=3. for ``bad'' galaxies. }
			\label{tab:cond_params}
			\end{centering}
		\end{table}

\section{Fake Data Testing for Ellipsoid Model}
\label{sec:fake_data_ellipsoid}

In this Appendix, we discuss how we tested our method for estimating the parameters of the velocity ellipsoid using fake data. 

To create fake data for a given line of sight for this model, we:

\begin{enumerate}
\item{Generate samples from our kernel density estimate for $M_{F81W}$ vs $M_{F606W}-M_{F814W}$ based on the weighted \cite{Vandenberg2006} isochrones.}
\item{Assign each draw an apparent magnitude, drawn from a uniform distribution in $m_{F814W}$ over the range $[19,24.5]$.}
\item{Given the resulting distances from the draws in apparent and absolute magnitudes, we use Monte Carlo acception/rejection to keep draws consistent with the MW density profile of \cite{Deason2011b}.}
\item{Assign stars velocities in spherical coordinates, based on random draws from normal distributions.}
\item{Convert $D, V_r, V_{\phi}, V_{\theta}$ to $\mu_l, \mu_b, v_{LOS}$ using the \verb+astropy.coordinates+ package. Given that we do not use \verb+astropy.coordinates+ to perform the velocity transformations in our ellipsoid modeling code, this step provides an additional check on our coordinate transformations.}
\item{Draw fake measured values from normal distributions centered on $\mu_l, \mu_b, v_{LOS}$, with dispersions corresponding to measurement uncertainties. For the purposes of this testing, we assign PM uncertainties of 0.2 mas yr$^{-1}$ and LOS velocity uncertainties based on a fit of the relation between apparent magnitude and LOS velocity error shown in Figure 7 of Paper I.}
\end{enumerate}

\begin{figure}
\includegraphics[width=\textwidth]{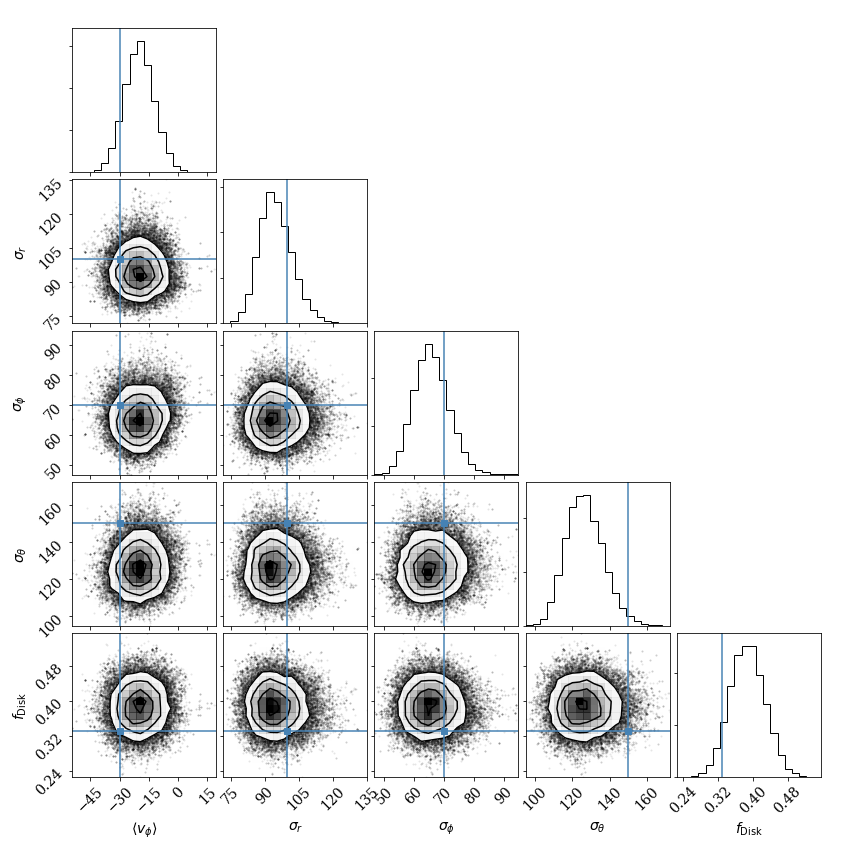}
\caption{Resulting projections of posterior samples for fake GOODS-N data. This fake sample contained 100 halo stars and 50 disk stars. The true values of the distributions used to generate the data are shown in blue.}
\label{fig:ellips_fake}
\end{figure}

We generate fake disk stars using a similar method, except drawing stars from the density profile and velocity distributions for our disk model. Figure \ref{fig:ellips_fake} shows the posterior distribution for the halo ellipsoid parameters when our analysis is performed on a fake dataset. This particular fake dataset contains 100 halo star and 50 disk stars in the GOODS-N field. Values for the parameters used to generate the data are shown in blue.

\begin{figure}
\centering
\includegraphics[width=\textwidth]{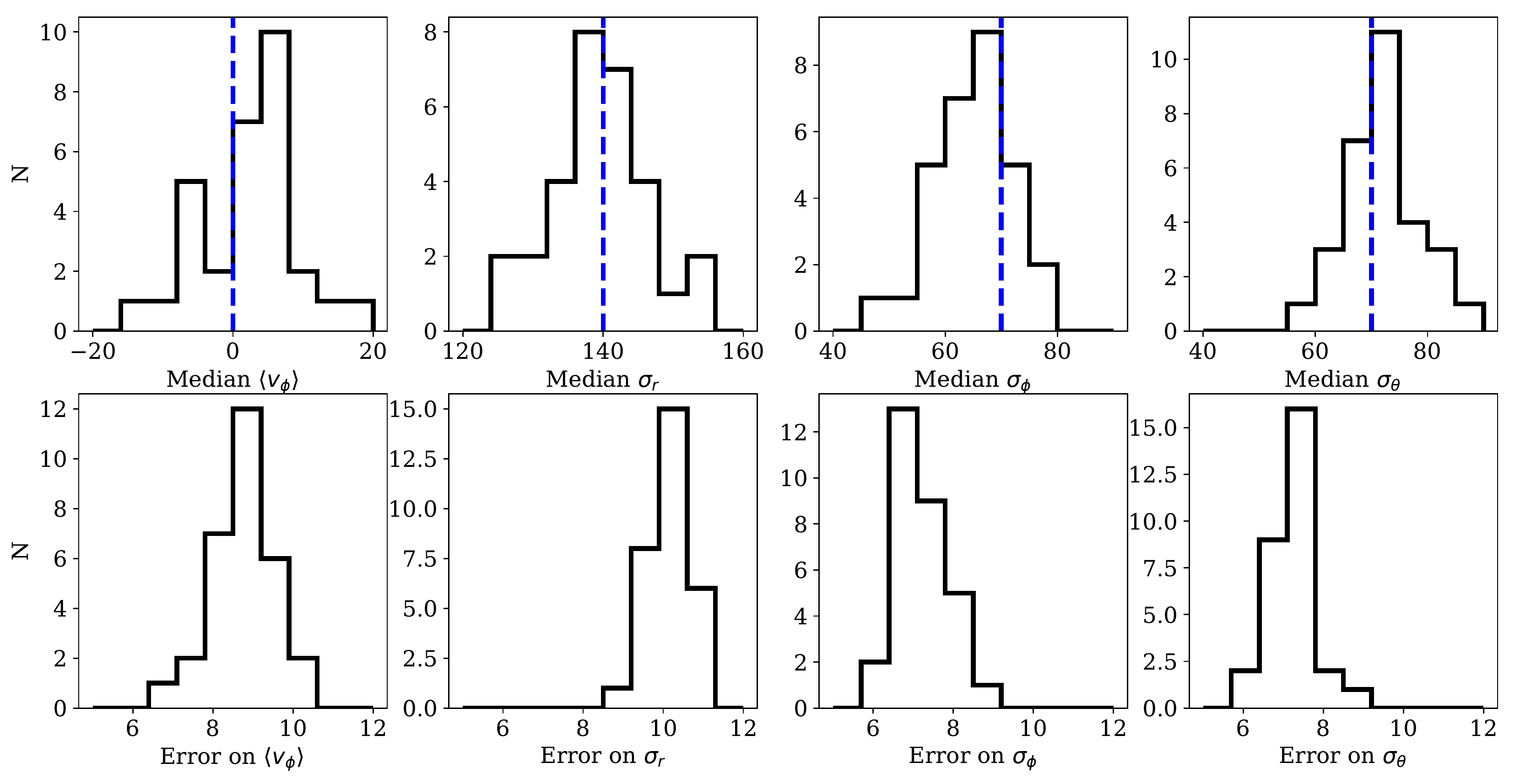}
\caption{Top panel: Distributions of posterior medians for the halo velocity ellipsoid parameters recovered from 30 fake datasets, each with 100 stars. Parameter values used to generate the fake data are shown as blue vertical dashed lines. Bottom panel: histograms of the the error estimates for each parameter.}
\label{fig:fake_100}
\end{figure}

Results from testing 30 fake halo datasets, each with 100 stars, are shown in Figure \ref{fig:fake_100}. Top panels show histograms of posterior medians for each simulated dataset; bottom panels are histograms of the errors measured in each dataset (computed as half the difference of the 84 and 16 percentiles). The errors in the posterior distributions are reasonable given the observed spread in posterior medians. The resulting distribution of posterior medians for $\beta$ are shown in the lefthand panel of Figure \ref{fig:beta_fake_goodss}.

\subsection{Sensitivity to Sample Size}

\begin{figure}
\centering
\includegraphics[width=0.47\textwidth]{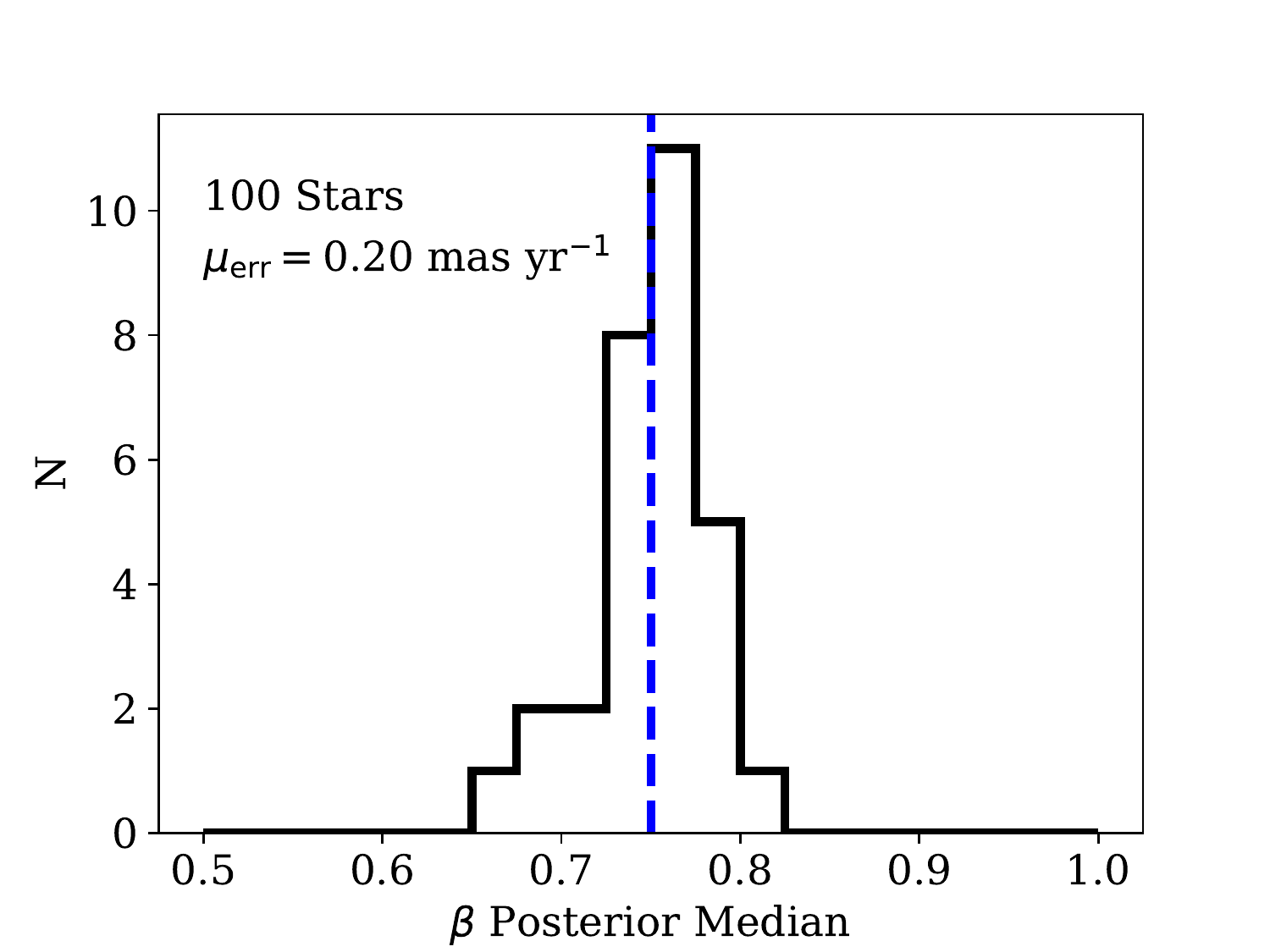}
\includegraphics[width=0.47\textwidth]{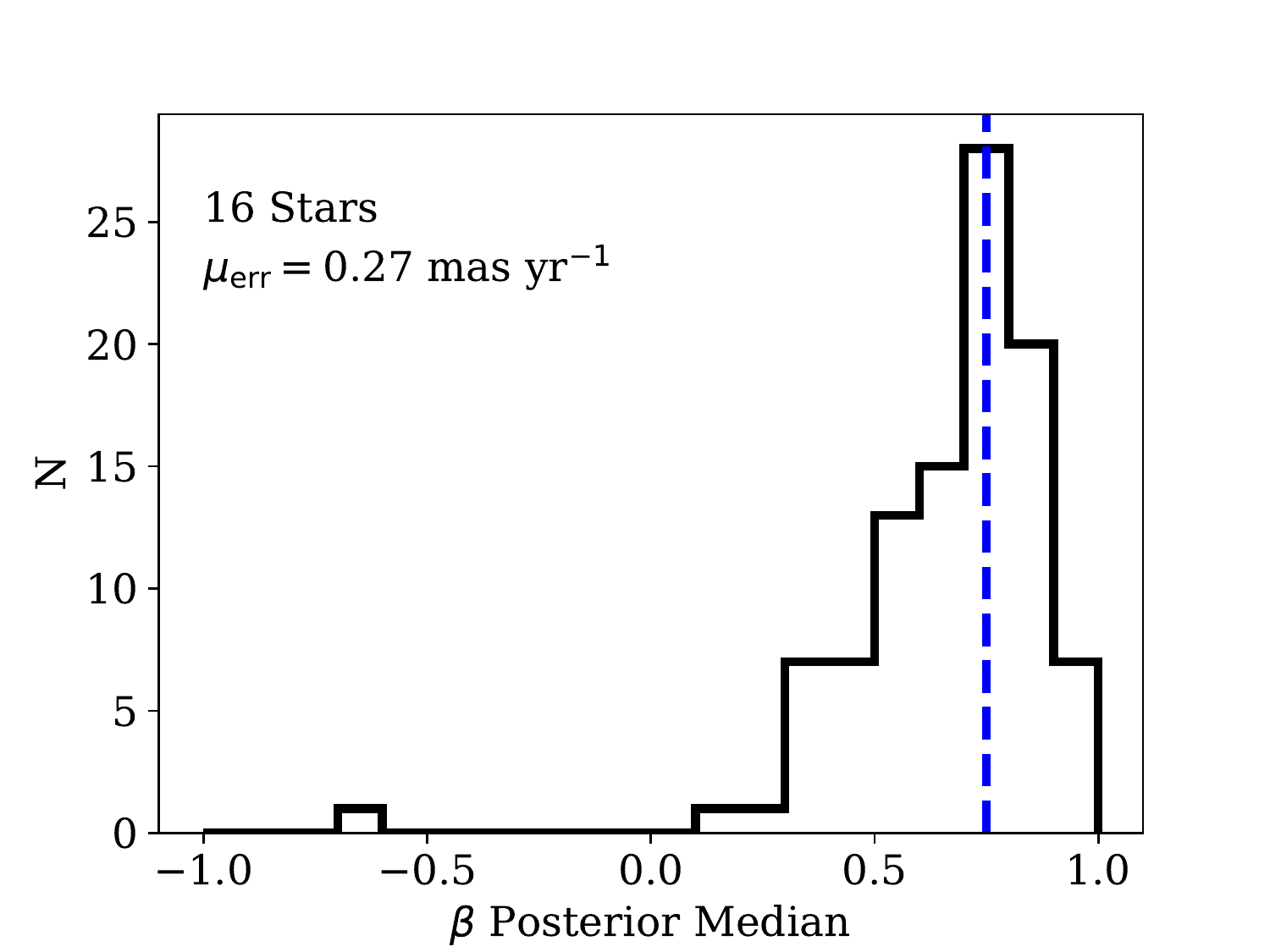}
\caption{Histograms of posterior medians for the estimates of $\beta$ from fake data testing. Lefthand panel: distribution of $\beta$ estimates from 30 fake datasets, each containing 100 stars, with PM uncertainties of 0.2 mas yr$^{-1}$. Righthand panel: the estimates of $\beta$ from 100 fake datasets, each containing 16 stars, with PM errors of 0.27 mas yr$^{-1}$. For both sets of fake datasets, radial velocity uncertainties were assigned as a function of apparent magnitude (see Figure 7 of Paper I). Only one out of the 100 fake datasets yielded a posterior median estimate of $\beta<0$.}
\label{fig:beta_fake_goodss}
\end{figure}

In order to assess how the sample size of the GOODS-S field is affecting the estimate of $\beta$ in that field, we generated 100 fake datasets, each containing 16 stars. These datasets were generated from velocity distributions that have $\beta_{\rm True}=0.75$. Figure \ref{fig:beta_fake_goodss} shows the distribution of the resulting posterior medians for $\beta$ when we model this fake dataset. Out of the 100 fake datasets, only one had posterior medians $\beta<0$. 

\end{document}